\newif\ifarxiv
\arxivfalse

\IfFileExists{fmcad2020.cfg}{\input{fmcad2020.cfg}}{}

\ifarxiv
  \documentclass{IEEEtran}
\else
  \documentclass{fmcad}
\fi

\usepackage[australian]{babel}
\usepackage[utf8]{inputenc}
\usepackage[final]{microtype} 
\usepackage[scaled=0.88]{helvet}
\usepackage{url}              
\usepackage[all]{foreign}     
\usepackage{booktabs}         
\usepackage{cite}
\usepackage[only,llbracket,rrbracket]{stmaryrd} 
\usepackage{amsmath,amssymb}  
\usepackage{thmtools}
\usepackage{empheq}
\usepackage{autonum}          
\usepackage{forest}           
\usepackage[inline]{enumitem}
\usepackage{listings}
\usepackage{lstautogobble}
\usepackage[export]{adjustbox}
\usepackage{wrapfig}
\usepackage{grffile}          
\usepackage{xcolor}
\usepackage{hyperref}
\usepackage[capitalise,nameinlink]{cleveref} 

\allowdisplaybreaks

\crefname{chapter}{Chap.}{Chaps.}
\crefname{section}{Sec.}{Secs.}
\crefname{appendix}{App.}{Apps.}
\crefname{equation}{Eq.}{Eqs.}
\crefname{figure}{Fig.}{Figs.}
\crefname{tabular}{Tab.}{Tabs.}
\crefname{definition}{Def.}{Defs.}
\crefname{proposition}{Prop.}{Props.}

\newtheorem{definition}{Definition}
\newtheorem{theorem}{Theorem}
\newtheorem{lemma}{Lemma}
\newtheorem{corollary}{Corollary}
\newtheorem{proposition}{Proposition}

\usepackage{packages/commands}
\usepackage{packages/graphics}
\usepackage{packages/symbols}

\graphicspath{ {./img/} }

\definecolor{primary}{RGB}{95,145,166}
\definecolor{secondary}{RGB}{242,130,48}
\colorlet{primary-bg}{primary!30}
\colorlet{secondary-bg}{secondary!30}

\providecommand\orcid[1]{}

\usepackage{lipsum}

\begin{document}
%
%
\title{%
  Reactive Synthesis from Extended \\ 
  Bounded Response LTL Specifications%
  \ifarxiv\thanks{%
    This paper has been accepted for publication in the \emph{Proceedings of the 2020 Formal Methods in Computer Aided Design conference}, FMCAD 2020,
    \url{https://fmcad.forsyte.at/}
  }\fi
}

\author{
  \IEEEauthorblockN{
    Alessandro Cimatti\IEEEauthorrefmark{1}
      \orcid{0000-0002-1315-6990},
    Luca Geatti\IEEEauthorrefmark{1}\IEEEauthorrefmark{2}
      \orcid{0000-0002-7125-787X},
    Nicola Gigante\IEEEauthorrefmark{2}
      \orcid{0000-0002-2254-4821},
    Angelo Montanari\IEEEauthorrefmark{2} 
      \orcid{0000-0002-4322-769X}
    and
    Stefano Tonetta\IEEEauthorrefmark{1}
      \orcid{0000-0001-9091-7899}
  }

  \IEEEauthorblockA{
    \IEEEauthorrefmark{1}
      Fondazione Bruno Kessler,
      Trento, Italy, \\
      Email: [cimatti,lgeatti,tonettas]@fbk.eu
  }
  
  \IEEEauthorblockA{
    \IEEEauthorrefmark{2}
      University of Udine,
      Udine, Italy, \\
      Email: [name.surname]@uniud.it
  }
}

\maketitle

\begin{abstract}
%
Reactive synthesis is a key technique for the design of correct-by-construction
systems and has been thoroughly investigated in the last decades. It consists
in the synthesis of a controller that reacts to environment's inputs satisfying
a given temporal logic specification. Common approaches are based on the
explicit construction of automata and on their determinization, which limit
their scalability.

In this paper, we introduce a new fragment of Linear Temporal Logic, called
Extended Bounded Response LTL (\LTLEBR), that allows one to combine bounded
 and universal unbounded temporal operators (thus covering
a large set of practical cases), and we show that reactive synthesis from
\LTLEBR specifications can be reduced to solving a safety game over
a deterministic symbolic automaton built directly from the specification. We
prove the correctness of the proposed approach and we successfully evaluate it
on various benchmarks.

\end{abstract}


\section{Introduction}
\label{sec:intro}
Since the dawn of computer science, synthesizing correct-by-construction
systems starting from a specification is
an important and difficult task.
A practical algorithm to solve this task would 
be
a big improvement in 
declarative programming, since it would allow the programmer
to write only the specification of the program, freeing her from possible
design or implementation errors, that, in many cases, are due to an imperative
style of programming.  In the context of formal verification and model-based
design, the possibility of synthesizing a controller able to comply with 
the specification for all possible behaviors of the environment
would be of great importance as well: all the effort would be directed to improve the
quality of the specification for the controller.

Reactive synthesis was first proposed by Church \cite{church1962logic} and
solved by B\"uchi and Landweber \cite{buchi1990solving} for \SoneS
specifications with an algorithm of nonelementary complexity. For Linear
Temporal Logic (\LTL) specifications, the problem has been shown to be
\EXPTIME[2]-complete \cite{pnueli1989synthesis,rosner1992modular}. In the attempt
of making reactive synthesis a practical task, in spite of its very high
complexity, research mainly focused on two lines:
\begin{enumerate*}[label=(\roman*)]
  \item finding good algorithms for the average case;
  \item restricting the expressiveness of the specification language.
\end{enumerate*}
Important examples of the first line of research are the contribution by Kupferman and
Vardi~\cite{kupferman2005safraless}, where the authors devise a procedure to
avoid Safra's determinization of B\"uchi automata (a known bottleneck in all the
problems requiring a determinization of a B\"uchi automaton), and the work by
Finkbeiner and Schewe~\cite{finkbeiner2013bounded}, where the problem is
reduced to a sequence of smaller problems on safety automata, obtained by
bounding the number of visits to a rejecting state of a co-B\"uchi automaton. A
meaningful example of restrictions to the specification language is the definition
of the \emph{Generalized Reactivity(1)} logic~\cite{piterman2006synthesis}, whose
synthesis problem can be solved in $\mathcal{O}(N^3)$ symbolic steps,
where $N$ is the size of the arena. Finally, in \cite{zhu2017symbolic} Zhu et al.\
consider reactive synthesis from Safety \LTL specifications. Although the
complexity remains doubly exponential, the proposed restriction allows one to reason on
finite words and thus to exploit
efficient tools for finite-state automata,
like, for instance, MONA \cite{henriksen1995mona}.

In this paper, we propose a new fragment of  \LTL, called \emph{Extended Bounded Response \LTL} (\LTLEBR for short), which supports \emph{bounded} operators
\cite{maler2007synthesizing}, such as $\ltl{G^{[a,b]}}$ and $\ltl{F^{[a,b]}}$,
along with universal unbounded temporal operators like $\ltl{G}$ and
$\ltl{R}$. We show that
formulas of \LTLEBR can be turned into
\emph{deterministic symbolic automata} over infinite words, with a translation
carried out in a completely symbolic way. 
Such a result is achieved in two steps:
(i) a \emph{pastification} of the subformulas containing only bounded
operators by making use of techniques similar to those
exploited for \MTL~\cite{maler2007synthesizing, maler2005real}, and (ii)
the construction of \emph{deterministic monitors} for the unbounded temporal operators. These two steps allow the entire
procedure to be carried out without ever producing any explicit automaton.
Then, we use existing algorithms for safety synthesis to solve the game on
the deterministic symbolic automaton. We implemented the proposed solution
in a tool,
called \tool, and compared its performance against state-of-the-art synthesizers
for full \LTL over a set of \LTLEBR formulas.
The outcomes of the experimental evaluation are
encouraging. 
\ifarxiv
  For lack of space, some of the proofs are reported in the appendix.
\else
  For lack of space, some of the proofs are reported in 
  \cite{CimattiGGMT20arxiv}.
\fi


\section{Preliminaries}
\label{sec:prel}
Linear Temporal Logic with Past (\LTLP) is a modal logic interpreted over
infinite state sequences. Let $\Sigma$ be a set of propositions. \LTLP formulas
are inductively defined as follows:\fitpar
  \begin{align}
    \phi \bydef p 
       &\choice \ltl{\lnot \phi}
       \choice \ltl{\phi_1 || \phi_2}
       \choice \ltl{X \phi} 
       \choice \ltl{\phi_1 U \phi_2}
       \choice \ltl{Y \phi} 
       \choice \ltl{\phi_1 S \phi_2}
  \end{align}
where $p \in \Sigma$. Temporal operators can be subdivided into the \emph{future
operators}, \emph{next} ($\ltl{X}$) and \emph{until} ($\ltl{U}$), and \emph{past
operators}, \emph{yesterday} ($\ltl{Y}$) and \emph{since} ($\ltl{S}$). We define
the following common abbreviations (where $\true$ stands for true):
\begin{enumerate*}[label=(\roman*)]
  \item $\ltl{X^i \phi}$ is $\ltl{X(X^{i-1}\phi)}$ if $i>0$ and $\ltl{X^0\phi}$ is
    $\phi$;
  \item \emph{release}: $\ltl{\phi_1 R \phi_2} \equiv \ltl{\lnot(\lnot\phi_1
    U \lnot\phi_2)}$;
  \item \emph{eventually}: $\ltl{F\phi_1} \equiv \ltl{\true U \phi_1}$;
  \item \emph{globally}: $\ltl{G\phi_1} \equiv \ltl{\lnot F \lnot\phi_1}$;
  \item \emph{trigger}: $\ltl{\phi_1 T \phi_2} \equiv \ltl{\lnot(\lnot\phi_1
    S \lnot\phi_2)}$;
  \item \emph{once}: $\ltl{O\phi_1} \equiv \ltl{\true S \phi_1}$;
  \item \emph{historically}: $\ltl{H\phi_1} \equiv \ltl{\lnot O \lnot\phi_1}$.
\end{enumerate*}

\LTL is obtained from \LTLP by allowing only the $\emph{next}$ and the
$\emph{until}$ operators. Conversely, \emph{Full Past \LTL} (\LTLFP) is the
fragment of \LTLP that only admits past operators. 

\LTL can also be enriched with \emph{bounded} temporal operators, such as the
\emph{bounded until} ($\ltl{\phi_1 U^{[a,b]} \phi_2}$) and \emph{bounded
eventually} ($\ltl{F^{[a,b]}\phi_1 \equiv \true U^{[a,b]}\phi_1}$). \emph{Full
Bounded \LTL} (\LTLFB) is the fragment of \LTL that includes only the
\emph{next}, \emph{bounded until}, and \emph{bounded eventually} operators. 

Let us now give the semantics of the above logics. A \emph{state sequence} is
an infinite sequence
$\sigma=\seq{\sigma_0,\sigma_1,\ldots}\in(2^\Sigma)^\omega$ of sets of
propositions $\sigma_i\in 2^\Sigma$, called \emph{states}. Given a sequence
$\sigma$, a position $i \ge 0$, and a formula $\phi$, the satisfaction of
$\phi$ by $\sigma$ at $i$, written $\sigma,i \models \phi$, is inductively
defined as follows:

\begin{tabular}{@{}lll@{}}
  $\sigma,i \models p$  & iff & $p \in \sigma_i$ \\
  $\sigma,i \models \lnot \phi$  & iff & 
    $\sigma,i \not \models \phi$ \\
  $\sigma,i \models \phi_1 \lor \phi_2$  & iff &
    either $\sigma,i \models \phi_1$ or
           $\sigma,i \models \phi_2$ \\
  $\sigma,i \models \phi_1 \land \phi_2$ & iff &
    $\sigma,i \models \phi_1$ and
    $\sigma,i \models \phi_2$ \\
  $\sigma,i \models \ltl{X\phi}$ & iff &
    $\sigma,i+1 \models \phi$ \\
  $\sigma,i \models \ltl{Y\phi}$ & iff &
    $i>0$ and $\sigma,i-1 \models \phi$ \\
  $\sigma,i \models \ltl{\phi_1 U \phi_2}$  & iff &
    there exists $j \ge i$ such that \\
    & & $\sigma,j \models \phi_2$ and $\sigma,k \models \phi_1$ for all \\
    & & $i \le k < j$ \\
  $\sigma,i \models \ltl{\phi_1 S \phi_2}$  & iff &
    there exists $j \le i$ such that \\ 
    & & $\sigma,j \models \phi_2$ and $\sigma,k \models \phi_1$ for all \\
    & & $j < k \le i$ \\
  $\sigma,i \models \ltl{\phi_1 U^{[a,b]} \phi_2}$  & iff &
    there exists $j \in [i+a,i+b]$ \\
    & & such that $\sigma,j \models \phi_2$ and \\ 
    & & $\sigma,k \models \phi_1$ for all $i \le k < j$ \\
\end{tabular}
We say that $\sigma$ satisfies $\phi$, written $\sigma \models \phi$, if and only if
$\sigma,0 \models \phi$.  We define the \emph{language} $\lang(\phi)$ of
a temporal formula $\phi$ as $\lang(\phi) = \{\sigma \in (2^\Sigma)^\omega \mid
\sigma \models \phi\}$. 

\subsection*{Symbolic safety automata and safety games}

To begin with, we formally define the problems of realizability and reactive synthesis for temporal formulas. 

As for realizability, it is convenient to view it as a two-player game between Controller, whose aim is to satisfy the specification,
and Environment, who tries to violate it.

\begin{definition}[Strategy]
  \label{def:strategy}
  Let $\Sigma=\Cset\cup\Uset$ be an alphabet partitioned into the set of 
  \emph{controllable} variables $\Cset$ and the set of \emph{uncontrollable} ones $\Uset$, such that $\Cset\cap\Uset=\emptyset$. A \emph{strategy for
  Controller} is a function $g:(2^\Uset)^+ \to 2^\Cset$ that, given the sequence
  $\Uncontr=\seq{\Uncontr_0,\ldots,\Uncontr_n}$ of choices made by \emph{Environment} so far,
  determines the current choices $\Contr_n=g(\Uncontr)$ of \emph{Controller}.
\end{definition}

Given a strategy $g:(2^\Uset)^+ \to 2^\Cset$ and an infinite sequence of
uncontrollable choices
$\Uncontr=\seq{\Uncontr_0,\Uncontr_1,\ldots}\in(2^\Uset)^\omega$, let
$g(\Uncontr)=\seq{\Uncontr_0\cup g(\seq{\Uncontr_0}), \Uncontr_1 \cup
g(\seq{\Uncontr_0, \Uncontr_1}), \ldots}$ be the state sequence resulting from
reacting to $\Uncontr$ according to $g$.

\begin{definition}[Realizability and Synthesis]
  \label{def:realizability}
  Let $\phi$ be a temporal formula over the alphabet $\Sigma = \Cset \cup
  \Uset$. We say that $\phi$ is \emph{realizable} if and only if there
  exists a strategy $g : (2^\Uset)^{+} \to 2^\Cset$ such that, for any
  infinite sequence $\Uncontr=\seq{\Uncontr_0, \Uncontr_1, \dots} \in
  (2^\Uset)^\omega$, it holds that $g(\Uncontr) \models \phi$. If $\phi$ is
  realizable, the synthesis problem is the problem of computing such a strategy $g$.
\end{definition}

Temporal logic has an intimate relationship
with automata on infinite words
\cite{vardi1994reasoning}, where different acceptance conditions give rise to
different 
classes of automata. For instance, 
the acceptance condition of (non-deterministic) Büchi automata allows them to recognize 
the class of $\omega$-regular
languages \cite{buchi1990decision}, including all languages definable by \LTLP
formulas. 

Here, we focus on a restricted type of acceptance condition, called \emph{safety} condition, and  we represent automata in a \emph{symbolic} way, as opposed to their common explicit representation. 

\begin{definition}[Symbolic Safety Automata]
  \label{def:ssa}
  A \emph{symbolic safety automaton} (SSA) is a tuple $\autom = (V, I, T, S)$, 
  where
  \begin{enumerate*}[label=(\roman*)]
    \item $V=X\cup\Sigma$, where $X$ is a set of \emph{state variables} and 
      $\Sigma$ is a set of \emph{input variables}, and
    \item $I(X)$, $T(X,\Sigma, X^\prime)$, and    $S(X)$, with 
      $X'=\set{x' \mid x\in X}$, are Boolean formulae which define the set of
      initial states, the transition relation, and the set of safe states,
      respectively.
  \end{enumerate*}
\end{definition}

In symbolic automata, states are identified by the values of state variables, and both
initial/final states and the transition relation are represented as Boolean
formulas. This allows them to be, in many cases, exponentially more succinct
than equivalent explicitly represented automata. In particular, the transition
relation $T(X,\Sigma,X')$ is built over state variables, input variables, and a
\emph{primed} version of state variables that represent the values of state
variables at the next state. As an example, if a variable $x$ has to flip at every
transition, the transition relation would contain a clause of the form $x \iff
\neg x'$.

\begin{definition}[Acceptance of SSA]
  \label{def:dssa-trace}
  Let $\autom$ be an SSA. A \emph{trace} is a sequence
  $\tau=\seq{\tau_0,\tau_1,\dots}\in(2^V)^\omega$ of subsets $\tau_i$ of
  $V$
  that satisfies the transition relation of $\autom$, that is, such
  that for all $i\ge 0$, $T(X,\Sigma,X')$ is satisfied when $\tau_i$ is used to
  interpret variables from $X$ and $\Sigma$, and $\tau_{i+1}$ is used to
  interpret variables from $X'$.
  We say that a trace $\tau$ is \emph{induced} by a word $\sigma
  = \seq{\sigma_0, \sigma_1, \dots} \in (2^\Sigma)^\omega$ iff $\sigma_i
  = \tau_i\cap\Sigma$ for all $i\ge 0$. 
  A trace $\tau$ is \emph{accepting} (or \emph{safe}) iff $\tau_i$ satisfies
  $S(X)$ for all $i \ge 0$. The \emph{language} of $\autom$, denoted as
  $\lang(\autom)$, is the set of all $\sigma \in (2^\Sigma)^\omega$ such that
  there exists an accepting trace induced by $\sigma$ in $\autom$.
\end{definition}

For reactive synthesis, a crucial property of an automaton $\autom$ is
\emph{determinism}, since in order to check if $\sigma \in \lang(\autom)$ it
suffices to check if \emph{the} trace induced by $\sigma$ in $\autom$ is
accepting.

\begin{definition}[Deterministic SSA]
\label{def:dssa}
  An SSA $\autom=(V,I,T,S)$ is \emph{deterministic} if:
  \begin{enumerate}
  \item the formula $I$ has exactly one satisfying assignment;
  \item the transition relation is of the form:
        \[T(X,\Sigma,X^\prime) \coloneqq 
          \bigwedge_{x \in X}(x^\prime \iff \beta_x(X\cup\Sigma))\]
  \end{enumerate}
  where each $\beta_x(X\cup\Sigma)$ is a Boolean formula over $X$ and $\Sigma$.
\end{definition}

Note that \cref{def:dssa} implies that for each $\sigma \in (2^\Sigma)^\omega$,
there exists exactly one trace induced by $\sigma$ for any given deterministic
SSA. The realizability and the synthesis problems can be defined over a
deterministic automaton as well; this gives rise to a safety game, which is
defined as follows.

\begin{definition}[Safety Game]
\label{def:safetygame}
  Let $\autom$ be a deterministic SSA over the alphabet $\Sigma = \Cset \cup
  \Uset$. A safety game is a tuple $G=\langle \autom, \Cset, \Uset \rangle$,
  where $\Cset$ and $\Uset$ are the sets of controllable and uncontrollable
  variables, respectively. We say that Controller wins the game if and only
  if there is a strategy $g : (2^\Uset)^{+} \to 2^\Cset$ such that for
  all sequences 
  $\Uncontr=\seq{\Uncontr_0, \Uncontr_1, \dots} \in (2^\Uncontr)^\omega$,
  \emph{the} trace $\tau$ induced by $g(\Uncontr)$ in $\autom$ is 
  \emph{accepting}.
\end{definition}


\section{Extended Bounded Response \LTL}
\label{sec:ebr-ltl}
In this section, we define \emph{Extended Bounded Response} \LTL, abbreviated \LTLEBR.
\LTLEBR extends \LTLFB (which  only features bounded operators) by admitting
Boolean combinations of the universal unbounded temporal operators 
\emph{release} ($\ltl{R}$) and \emph{globally} ($\ltl{G}$).

\begin{definition}[The logic \LTLEBR]
\label{def:ltlebr}
  Let $a,b \in \N$. An
  \LTLEBR formula $\chi$ is
  inductively defined as follows:
  \begin{align}
    \psi \bydef p 
       &\choice \ltl{\lnot \psi}
       \choice \ltl{\psi_1 || \psi_2}
       \choice \ltl{X \psi} 
       \choice \ltl{\psi_1 U^{[a,b]}\psi_2} &
          \text{\footnotesize Full Bounded Layer} \\
    \phi \bydef \psi
       &\choice \ltl{\phi_1 \land \phi_2}
       \choice \ltl{X \phi}
       \choice \ltl{G \phi} 
       \choice \ltl{\psi R \phi} &
          \text{\footnotesize Future Layer} \\
    \chi \bydef \phi 
      &\choice \ltl{\chi_1 \lor \chi_2}
      \choice \ltl{\chi_1 \land \chi_2} & 
          \text{\footnotesize Boolean Layer}
  \end{align}
\end{definition}

We refer to \cref{sec:prel} for the semantics of
\LTLEBR operators. In the next sections, we will show how 
to build, given an \LTLEBR formula $\phi$, a deterministic symbolic safety automaton
$\autom(\phi)$ such that $\lang(\autom(\phi)) = \lang(\phi)$.

\subsection{Examples}
We now give some simple examples of requirements that can be expressed in the \LTLEBR logic.

The first one is a typical bounded response requirement: Controller
has to answer a grant $g$ at most $k$ time units after the request $r$ of
Environment is issued. It can be expressed by the following \LTLEBR formula:
\begin{align}
  \ltl{G(r \to F^{[0,k]}g)}
\end{align}

Another quite common requirement is \emph{mutual exclusion}. As an example, the
case of an arbiter that has to grant a resource to at most one client at once
can be captured as follows (for each $i$, $g_i$ means that the resource has been
granted to client $i$):\fitpar
\begin{align}
  \ltl{G(\bigwedge_{1 \le i < j \le n}\lnot (g_i \land g_j))}
\end{align}

When a set of clients with different priorities has to be managed, it is possible to introduce a requirement stating that, whenever two or more clients simultaneously send a request, clients with a higher priority must be granted before those with a lower one ($i < j$ means that the priority of client $i$ is higher than that of client $j$):
\begin{align}
  \ltl{\bigwedge_{1 \le i < j \le n} G((r_i \land r_j) \to (\lnot g_j) U^{[0,k]} g_i)}
\end{align}

Finally, in many situations it is important to include requirements about the
\emph{configuration} of a system model. 
Consider the case of a thermostat. One may ask that if the \texttt{prog}
modality is off, then the controller has to communicate the signal \texttt{on}
to the boiler for an indefinitely long amount of time, while, in case the
\texttt{prog} modality is on, it has to do that only for a specific interval of
time, say $[h_1,h_2]$, after which it has to stop the communication with the boiler.  This can be expressed in \LTLEBR by the following formula:
\begin{align}
  \ltl{
    (\lnot \texttt{prog} \land G(\texttt{on})) 
      \lor
    (\texttt{prog} \land G^{[h_1,h_2]}(\texttt{on}) \land X^{h_2}G(\texttt{off}))
  }
\end{align}

\subsection{Comparison with other temporal logics}
\label{sub:deltaexpress}
Zhu \emph{et al.}~\cite{zhu2017symbolic} studied the synthesis problem for 
\emph{Safety \LTL}, which can be viewed as the \emph{until}-free fragment of \LTL
in negated normal form (NNF). Every formula $\phi$ of \LTLEBR can be turned into
a Safety \LTL one by
\begin{enumerate*}[label=(\roman*)]
  \item transforming $\phi$ in NNF and
  \item expanding each bounded operator in terms of conjunctions or 
        disjunctions.
\end{enumerate*}
As an example, the \LTLEBR formula $\phi \coloneqq \ltl{G(p \to F^{[0,5]}q)}$
is equivalent to the Safety \LTL formula $\phi^\prime \coloneqq \ltl{G(p \to
\bigvee_{i=0}^{5}X^i q)}$. However, since constants in \LTLEBR are represented
by using a logarithmic encoding, \LTLEBR formulas can be exponentially more
succinct than Safety \LTL ones. Whether the converse holds as well, \ie whether
any formula of Safety \LTL can be translated into an equivalent \LTLEBR one, is
still an open question. As an example, $\ltl{G(p \lor G q)}$ is a Safety \LTL
formula but, syntactically, is not an \LTLEBR one.

Maler \emph{et al.}~\cite{maler2007synthesizing} introduced  \emph{Metric
Temporal Logic with a Bounded-Horizon} (\MTLB for short) as the metric temporal
logic with \emph{only} bounded operators interpreted over dense time. They
addressed the problem of reactive synthesis from \MTLB specifications by
showing that each \MTLB formula can be transformed into a \emph{deterministic}
timed automaton. With respect to this fragment, and ignoring the differences in
the underlying temporal structures (in our setting, time is discrete), \LTLEBR
extends \MTLB with Boolean combinations of unbounded universal temporal
operators.


\section{%
  From \LTLEBR to\texorpdfstring{\\}{ }deterministic symbolic safety automata%
}
\label{sec:ebr-ltl-to-dssa}
This section focuses on the procedure to turn every \LTLEBR formula into
a deterministic symbolic safety automaton on infinite words (see
\cref{def:dssa}) that recognizes the same language. 

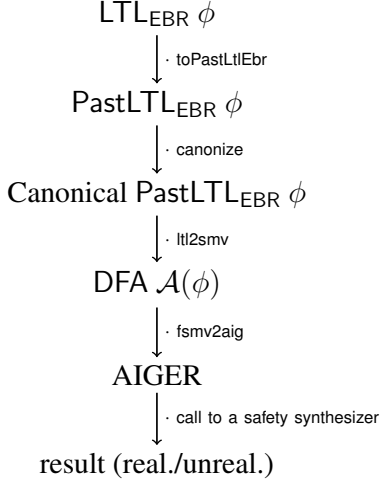
\begin{figure}
\centering
\begin{adjustbox}{minipage=\linewidth,scale=.8}
    \begin{tikzpicture}[
    ph_name/.style={font=\Large},
    subph_name/.style={font=\large},
    ph_arrow/.style={thick, ->},
    bounds/.style={font=\small},
    ]
    \node[ph_name](ph1) at (1.0,8.5) {$\LTLEBR \ \phi$};
    \node[ph_name](ph2) at (1.0,7.0) {$\PLTLEBR \ \phi$};
    \draw[ph_arrow] (ph1) edge
        node[bounds,right,text width=3cm,align=left] {
        $\cdot$ \toPastEbr
        } (ph2);
    \node[ph_name](ph3) at (1.0,5.5) {Canonical \PLTLEBR $\phi$};
    \draw[ph_arrow] (ph2) edge
        node[bounds,right,text width=3cm,align=left] {
        $\cdot$ \canonize
        } (ph3);
    \node[ph_name](ph4) at (1.0,4.0) {\DFA $\autom(\phi)$};
    \draw[ph_arrow] (ph3) edge 
        node[bounds,right] {
        $\cdot$ \textsf{ltl2smv}
        } (ph4);
    \node[ph_name](ph5) at (1.0,2.5) {AIGER};
    \draw[ph_arrow] (ph4) edge 
        node[bounds,right] {
        $\cdot$ \textsf{fsmv2aig}
        } (ph5);
    \node[ph_name](ph6) at (1.0,1.0) {result (real./unreal.)};
    \draw[ph_arrow] (ph5) edge 
        node[bounds,right,text width=5cm] {
        $\cdot$ \textsf{call to a safety synthesizer}
        } (ph6);
    \end{tikzpicture}
    \caption{The overall procedure.}
    \label{fig:main}
\end{adjustbox}
\end{figure}

In doing that, we apply a few transformation steps on the formula, summarized in
\cref{fig:main}, to simplify its syntactic structure and turn it into a form
amenable to direct transformation into a deterministic SSA. We define two
syntactic restrictions of \LTLEBR that are the targets of the transformation
steps.
\begin{definition}[\PLTLEBR]
\label{def:past_ebr}
  An \PLTLEBR formula $\chi$ is inductively defined as follows:
  \begin{align}
    \psi \bydef p 
        &\choice \ltl{\lnot \psi}
        \choice \ltl{\psi_1 || \psi_2}
        \choice \ltl{Y \psi} 
        \choice \ltl{\psi_1 S \psi_2} \\
    \phi \bydef \psi
        &\choice \ltl{\phi_1 \land \phi_2}
        \choice \ltl{X \phi}
        \choice \ltl{G \phi} 
        \choice \ltl{(X^i \psi) R \phi} \\
    \chi \bydef \phi 
      &\choice \ltl{\chi_1 \lor \chi_2}
      \choice \ltl{\chi_1 \land \chi_2}
  \end{align}
\end{definition}

\begin{definition}[Canonical \PLTLEBR]
\label{def:canonical_ebr}
  The \emph{canonical form} of \PLTLEBR formulas is inductively defined as follows:
  \begin{align}
    \psi \bydef p 
        &\choice \ltl{\lnot \psi}
        \choice \ltl{\psi_1 || \psi_2}
        \choice \ltl{Y \psi} 
        \choice \ltl{\psi_1 S \psi_2} \\ 
    \phi \bydef \psi
        &\choice \ltl{G \psi} 
        \choice \ltl{\psi_1 R \psi_2} \\
    \lambda \bydef \phi
      &\choice \ltl{X \lambda} \\
    \chi \bydef \lambda
      &\choice \ltl{\chi_1 || \chi_2}
      \choice \ltl{\chi_1 && \chi_2} 
  \end{align}
\end{definition}

Canonical \PLTLEBR formulas do not contain nested occurrences of unbounded
temporal operators, whose operands can be only full-past formulas, and each of
these is prefixed by an arbitrary number of \emph{next} operators.

The transformation of \LTLEBR formulas into deterministic SSAs consists of three
steps:
\begin{enumerate*}[label=(\roman*)]
  \item a translation from \LTLEBR to \PLTLEBR;
  \item a translation from \PLTLEBR to its canonical form;
  \item a transformation of canonical \PLTLEBR formulas into deterministic SSAs.
\end{enumerate*}
Once a deterministic SSA $\autom(\phi)$ for the original \LTLEBR formula $\phi$ over $\Cset \cup \Uset$ has been obtained,  
to solve the safety game $\langle \autom(\phi), \Cset, \Uset \rangle$, \ie, to decide the existence of
a strategy for Controller in the automaton, we apply an existing safety synthesis
algorithm (see \cref{def:safetygame}).


\subsection{From \LTLEBR to \PLTLEBR}
\label{sub:pastify}
Let $\phi$ be an \LTLEBR formula. The first step consists in translating each
\LTLFB subformula of $\phi$ into an \emph{equivalent} one, which is of the form
$X^d\psi$, with $\psi \in \LTLFP$ and $d \in \N$. We refer to this process as
\emph{pastification} \cite{maler2007synthesizing, maler2005real}. As we will see, since ``the past has already happened", full-past formulas can be represented by deterministic monitors.

In order to pastify each \LTLFB subformula of $\phi$, we adapt to \LTLEBR a technique developed by Maler \emph{et
al.} for \MTLB~\cite{maler2007synthesizing, maler2005real}. Intuitively, for each model of a full-bounded formula $\phi$, there exists a furthermost time point $d$ (the \emph{temporal depth}
of $\phi$) such that
the subsequent states cannot be constrained by $\phi$ in any way. The \emph{pastification} of $\phi$ is a formula that uses only past
operators and that is equivalent to $\phi$ when interpreted at time point
$d$ instead of at the origin.

\begin{definition}[Temporal Depth \cite{maler2007synthesizing}]
\label{def:tempdepth}
  Let $\phi$ be an \LTLFB formula. The \emph{temporal depth} of
  $\phi$, denoted as $D(\phi)$, is inductively defined as follows:
  \begin{itemize}
    \item $D(p) = 0$, for all $p \in \Sigma$
    \item $D(\lnot\phi_1) = D(\phi_1)$
    \item $D(\phi_1 \land \phi_2) = \max\{D(\phi_1),D(\phi_2)\}$
    \item $D(\ltl{X\phi_1}) = 1 + D(\phi_1)$
    \item $D(\ltl{\phi_1 U^{[a,b]} \phi_2}) = b +\max\{D(\phi_1),D(\phi_2)\}$
  \end{itemize}
\end{definition}
Let $M_\phi$ (only $M$ if unambiguous) be the greatest constant in $\phi$, with
$M_\phi=0$ if $\phi$ has no constants. It can be observed that $D(\phi)\le
M\cdot n$, where $n=|\phi|$.

\begin{definition}[Pastification \cite{maler2007synthesizing}]
  \label{def:pastification}
  Let $\phi$ be an \LTLFB formula and $d\ge D(\phi)$. The pastification of 
  $\phi$ is the formula $\Pi(\phi,d)$ inductively defined as follows:
  \begin{itemize}
    \item 
      $\Pi(p,d) = \ltl{Y^d p}$ 
    \item
      $\Pi(\lnot\phi,d) = \lnot \Pi(\phi,d)$
    \item
      $\Pi(\phi_1 \land \phi_2,d) = \Pi(\phi_1,d) \land \Pi(\phi_2,d)$
    \item
      $\Pi(\ltl{X\phi},d) = \ltl{\Pi(\phi,d-1)}$
    \item
      $\Pi(\ltl{\phi_1 U^{[a,b]} \phi_2},d)
      = \\
        \quad\bigvee_{t=0}^{b-a}(\ltl{Y^t(\Pi(\phi_2, d-b)} \land 
          \ltl{H^{b-t-1}Y\Pi(\phi_1,d-b))})$
  \end{itemize}
\end{definition}
Note that from \cref{def:pastification} we can derive that
$\ltl{\Pi(F^{[a,b]}\phi,d)==\Pi(\top
U^{[a,b]}\phi,d)==\bigvee_{t=0}^{b-a}Y^t\Pi(\phi,d-b) }$, which can be
succinctly written using the \emph{once} operator, hence we can define
$\ltl{\Pi(F^{[a,b]}\phi,d)=O^{[0,b-a]}\Pi(\phi,d-b)}$.
\begin{proposition}[Soundness of pastification]
  \label{prop:pastification}
  Let $\varphi$ be a \LTLFB formula. For all state sequences
  $\sigma \in (2^\Sigma)^\omega$, all $i \in \N$, and all $d\ge D(\phi)$, 
  it holds that:
  \begin{align}
    \sigma,i \models \varphi \ &\Leftrightarrow \ 
    \sigma,i \models
    \ltl{X^d \Pi(\varphi,d)}
  \end{align}
\end{proposition}

From now on, let $\pastify(\phi)$ be the formula
$X^{D(\phi)}\Pi(\phi,D(\phi))$.  As an example, if $\phi \coloneqq \ltl{F^{[0,
k_1]}(q \land F^{[0,k_2]} p)}$, then $\pastify(\phi) \coloneqq
X^{k_1+k_2}\ltl{O^{[0,k_1]} (Y^{k_2}q \land O^{[0,k_2]}p)}$.  We state the
following complexity result about pastification.

\begin{proposition}
\label{prop:pastsize}
  Let $\phi$ be a \LTLFB formula. Then, $\pastify(\phi)$ is a formula of size
  $\mathcal{O}(n^2 \cdot M^{\log_2 n + 1})$, where $n = |\phi|$ and $M$ is
  the greatest constant in $\phi$.
\end{proposition}
\ifarxiv
  \begin{IEEEproof}
  See the appendix.
  \end{IEEEproof}
\fi

Note that if $\phi$ has no constants, that is, $M=1$, the size of $\pastify(\phi)$ is $\mathcal{O}(n^2)$ . Given an \LTLEBR formula $\phi$, we pastify each of its
\LTLFB subformulas with the \pastify operator: we call this step \toPastEbr.
Once it has been completed, the resulting formula belongs to \PLTLEBR. 

The \toPastEbr algorithm can be improved by observing that there are \LTLFB formulas that already belong to \PLTLEBR. One example is the formula $\ltl{p \land XXXq}$. Obviously, for this kind of formulas there is no need for the algorithm to pastify them. Consider the previous example. Without the proposed trick, the algorithm would have produced the formula $\ltl{XXX(YYYp
\land q)}$, while, by simply noticing that the formula already belongs to \PLTLEBR, it does not need to pastify anything, returning \ltl{p \land XXXq}.

\begin{proposition}
\label{prop:pastebrsize}
  For each \LTLEBR formula $\phi$, there is an equivalent \PLTLEBR formula
  $\phi^\prime$ of size $\mathcal{O}(n^3 \cdot M^{\log_2 n + 1})$, where
  $n = |\phi|$ and $M$ is the greatest constant in $\phi$.
\end{proposition}
\begin{IEEEproof}
  Let $\phi$ be an \LTLEBR formula and let $\phi^\prime \coloneqq
  \toPastEbr(\phi)$. By \cref{prop:pastification}, the \toPastEbr algorithm replaces the \LTLFB subformulas of $\phi$ with an equivalent formula, hence  $\phi \equiv \phi^\prime$. Since in $\phi$ there are at most $n=|\phi|$
  subformulas, then, by \cref{prop:pastsize}, $|\phi^\prime| = n \cdot
  \mathcal{O}(n^2 \cdot M^{\log_2 n + 1})$, that is, $|\phi^\prime|
  = \mathcal{O}(n^3 \cdot M^{\log_2 n})$.
\end{IEEEproof}
Note that if there are no constants in $\phi$, that is, $M=1$, then,  by \cref{prop:pastsize}, $|\toPastEbr(\phi)| = \mathcal{O}(n^3)$.


\subsection{From \PLTLEBR to Canonical \PLTLEBR}
\label{sub:canonize}
The second step is the canonization of the \PLTLEBR formula obtained from the
previous step, in order to obtain an equivalent formula in canonical form
(\cref{def:canonical_ebr}). Canonical \PLTLEBR formulas are Boolean
combinations of formulas of the form $\ltl{X^i\psi_1}$, $\ltl{\ltl{X^iG
\psi_1}}$, and $\ltl{X^i(\psi_1 R \psi_2)}$, where $\psi_1$ and $\psi_2$ are
full past formulas.  Compared to general \PLTLEBR formulas, formulas in
canonical form do not admit neither nested unbounded operators nor \emph{next}
operators in front of the left-hand argument of a \emph{release}. The
canonization of a \PLTLEBR formula is obtained by applying a set of rewriting
rules.

\begin{definition}[Canonization]
\label{def:canonize}
  Given a \PLTLEBR formula $\phi$, $\canonize(\phi)$ is the formula obtained by
  recursively applying the $R_1$-$R_7$ rules to the subformulas of $\phi$ in
  a bottom-up fashion followed by the application of the $R_{flat}$ rule: 
  \begin{align}
    &R_1: \ltl{X(\psi_1 \land \psi_2)       \leadsto 
      X\psi_1 \land X\psi_2 } \\
    &R_2: \ltl{\psi R (\psi_1 \land \psi_2) \leadsto 
      \psi R \psi_1 \land \psi R \psi_2 }\\
    &R_3: \ltl{(X^i \psi_1) R (X^j \psi_2)}  \leadsto \\ 
      &\qquad\begin{cases}
        \ltl{X^i(\psi_1 R (Y^{i-j}\psi_2))}
          & \mbox{ if } i>j \\ 
        \ltl{X^j((Y^{j-i}\psi_1) R \psi_2)} 
          & \mbox{ otherwise}
      \end{cases} \\
    &R_4: \ltl{(X^i \psi_1)R(X^j(\psi_2 R \psi_3))} \leadsto \\
      &\qquad\begin{cases}
        \ltl{X^i( \psi_1 R ((Y^{i-j}\psi_2) R (Y^{i-j} \psi_3) ) )}
          &\mbox{ if } i>j \\
        \ltl{X^j( (Y^{j-i}\psi_1 ) R (\psi_2 R \psi_3) )} 
          &\mbox{ otherwise }
      \end{cases} \\
    &R_5: \ltl{GX^iG\psi} \leadsto
      \ltl{X^iG\psi} \\
    &R_6: \ltl{GX^i(\psi_1 R \psi_2)} \leadsto
      \ltl{X^i G \psi_2} \\
    &R_7: \ltl{(X^i\psi_1) R (X^j G \psi_2 )} \leadsto \\
      &\qquad\begin{cases}
        \ltl{X^i G Y^{i-j} \psi_2}
          &\mbox{ if } i > j \\
        \ltl{X^j G \psi_2}
          &\mbox{otherwise}
      \end{cases} \\
    &R_{flat}: \ltl{X^i(\psi_1 R (\psi_2 R (\dots (\psi_{n-1} R \psi_n)\dots)))}
      \leadsto \\
      &\qquad\ltl{X^i( (\psi_{n-1} \land O(\psi_{n-2} \land \dots O(\psi_1
      \land Y^i \true)\dots)) R \psi_n )} \\
      &\qquad \mbox{for any } n \ge 3
    \end{align}
  where $\psi$, $\psi_1$, $\psi_2$, and $\psi_3$ are full-past formulae.
\end{definition}

It is worth noticing that, as far as for now, we do not have rules (preserving
the equivalence) to deal with the following cases: 
\begin{enumerate*}[label=(\roman*)]
  \item
    $\ltl{(\phi_1 \land \phi_2) R (\phi)}$,
  \item
    $\ltl{(G\phi_1)R(\phi)}$ or 
  \item
    $\ltl{(\phi_1 R \phi_2) R (\phi)}$.
\end{enumerate*}
This is why in \cref{def:ltlebr} we restricted the left-hand argument of each
\emph{release} operator to be a full-bounded formula.

\begin{lemma}[Soundness of $\canonize(\cdot)$]
  For any \PLTLEBR formula $\phi$, it holds that $\phi$ and $\canonize(\phi)$
  are equivalent and $\canonize(\phi)$ is a Canonical \PLTLEBR formula.
\end{lemma}
\ifarxiv
  \begin{IEEEproof}
  See the appendix.
  \end{IEEEproof}
\fi

\begin{proposition}[Complexity of $\canonize(\cdot)$]
\label{prop:canonizesize}
  For any \PLTLEBR formula $\phi$, $\canonize(\phi)$ can be built in
  $\mathcal{O}(n)$ time, and the size of $\canonize(\phi)$ is $\mathcal{O}(n)$, 
  where $n = |\phi|$.
\end{proposition}
\ifarxiv
  \begin{IEEEproof}
  See the appendix.
  \end{IEEEproof}
\fi

\subsection{From Canonical \PLTLEBR to deterministic SSA}
\label{sub:dssa}
The particular shape of canonical \PLTLEBR formulas 
makes it possible to
encode the specification into deterministic SSAs. The key observation is that \LTLFP formulas can be encoded into deterministic automata: since these formulas talk exclusively about the past, their truth can be evaluated  at any single step depending only on previous steps, without making any guess about the future (``the past already happened''). But \LTLFP formulae are not
the only ones that can be encoded deterministically. Consider, for instance, the formula $\phi\equiv\ltl{Xp \lor Xq}$. At a first glance, it may seem that $\phi$
needs a non-deterministic automaton to be encoded, which at the first state makes
a choice about whether $p$ or $q$ will hold in the next state. Nevertheless,
this formula is equivalent to $X(p \lor q)$ and it corresponds to the
\emph{deterministic} automaton that, once arrived in its second state by reading
any proposition symbol, proceeds to an accepting state by reading either $p$ or
$q$, or goes to a sink (\emph{error}) state otherwise.

\PLTLEBR in its canonical form combines full past formulas into a broader
language that can still be turned into symbolic deterministic automata,
extending the above intuition and exploiting the \emph{monitorability} of
\emph{universal} temporal operators.

Monitoring is a technique coming from 
\emph{runtime verification}~\cite{leucker2009brief}. Consider the formula
$\ltl{G\alpha}$. By observing a state sequence, at each step we can decide if a
\emph{violation} has occurred; indeed, if $\alpha$ is false at the current
step, then the value of $\ltl{G\alpha}$ is certainly false for each of the
previous steps. More generally, universal temporal formulas, such as
$\ltl{G\phi}$ and $\ltl{\phi_1 R\phi_2}$, are \emph{monitorable}, meaning that a
violation of them can be decided on the basis of the observation of a \emph{finite}
number of steps. In particular, reporting an error in the next state can be done
by considering only the current values. This means that any universal temporal
operator can be monitored by adding a Boolean \emph{error variable} with a
\emph{deterministic} transition relation.

Therefore, despite not being able to evaluate the truth of a formula such as
$\ltl{G\alpha}$, as it can be done in the case of past operators, we can nevertheless state in the accepting condition that an error state can never be reached. In this way, if the trace is accepting, that is, an error state can never be reached, then we know that there are no violations, \eg for $\ltl{G\alpha}$, we have forced $\alpha$
to be true in every state. Otherwise, if the trace is not accepting, that is, an
error state is reachable, we know that there is a (finite) violation and that
the temporal formula was falsified at some step. We therefore introduce an
\emph{error bit} for each $\ltl{X^i \psi_1}$, $\ltl{X^i G\psi_1}$, and
$\ltl{X^i(\psi_1 R \psi_2)}$ of a canonical \PLTLEBR formula. 


Let $\phi$ be a canonical \PLTLEBR formula over the alphabet $\Sigma = \Cset
\cup \Uset$.  We define the deterministic SSA $\autom(\phi) = (V,I,T,S)$ as
follows:
\begin{itemize}
  \item \emph{Variables.} The set of \emph{state variables} of the automaton is defined as $X = X_P \cup X_F \cup X_C$, where:
  \begin{align}
    X_P & = \set{v_\alpha\mid\text{$\alpha$ is an \LTLFP subformula of $\phi$}}\\
    X_F & = \set*{error_\varphi \,\middle|\, 
      \begin{aligned}
        &\text{$\varphi$ is subformula of $\phi$ of the form}\\
        &\text{%
          $\ltl{X^i\psi}$, $\ltl{X^iG\psi}$, or 
          $\ltl{X^i(\psi_1 R \psi_2)}$%
        }
      \end{aligned}} \\
    X_C & = \set*{counter_i \,\middle|\, \begin{aligned}
      & i \in \set{0, \dots, \log_2 d}\\
      &\text{%
        $d$ max. among all $\ltl{X^d\psi}$ in $\phi$.
      }
    \end{aligned}}
  \end{align}
  Intuitively, variables in $X_P$ track the truth value of all the full-past
    subformulas, variables in $X_F$ implement the above-described monitoring mechanism, and variables in $X_C$ are used to encode a binary
    counter used to monitor nested \emph{tomorrow} operators. In particular,
    for $n$ nested \emph{tomorrow} operators, a counter with $\log_2(n)$
    bits is needed.
      
  \item \emph{Initial state.} 
    All the state variables, including the counter bits, are initially false, that is, $I(X) = \bigwedge_{x \in X} \lnot x$.

  \item \emph{Transition relation.} 
    $T(X,\Sigma,X^\prime)$ is the conjunction of the transition
    \emph{functions} of the binary counter and the monitors of each subformula
    of $\phi$, as will be defined later. Notice that each conjunct is of the
    form $x^\prime \iff \beta(X \cup \Sigma)$, and thus it is a deterministic
    transition relation.

  \item \emph{Safety condition.}
    $S(X)$ is a Boolean formula obtained from $\phi$
    by replacing  each formula $\varphi\in X_F$ by $\lnot error_\varphi$, \ie
    $S(X) = \phi[\varphi / \lnot error_\varphi]$.
\end{itemize}

We now define the monitors for the binary counter, used to handle nested
\emph{tomorrow} operators, any formula $\psi\in\LTLFP$, and any canonical \PLTLEBR formula of one of the forms $\ltl{X^i \psi_1}$, $\ltl{X^i G\psi_1}$, and
$\ltl{X^i(\psi_1 R \psi_2)}$. We give the definition of the monitors using the
SMV language \cite{cavada2014nuxmv}, as it provides useful shorthands (like
the \emph{switch-case} primitive). Each of the following SMV statement
corresponds to the Boolean formula that defines transition functions of our
monitors.

\lstset{
  mathescape=true,
  frame=bt,
  basicstyle=\scriptsize,
  autogobble=true
}

The monitor for the counter is defined as follows:
\begin{lstlisting}{Name}
  next(counter$_0$) := $\lnot$ counter$_0$
  next(counter$_i$) := (counter$_{i-1}$ $\lor$ counter$_i$) $\land$ $\lnot$counter$_i$
\end{lstlisting}

If $\psi \coloneqq \ltl{\alpha S \beta}$ or $\ltl{Y\alpha}$,
its monitor is defined as follows:
\begin{lstlisting}{Name}
  next($v_{\ltl{Y \alpha}}$) := $v_{\alpha} \land {}$counter$>0$
  DEFINE 
    $v_{\ltl{\alpha S \beta}}$ := $v_\beta \lor (v_\alpha \land v_{\ltl{Y(\alpha
    S \beta)}})$
\end{lstlisting}

If $\psi$ is a propositional atom, a negation, or a disjunction of full-past formulas, we define its monitor 
as follows:
\begin{lstlisting}{Name}
  DEFINE
    $v_{p}$ := $p$
    $v_{\lnot\alpha}$ := $\lnot v_{\alpha}$
    $v_{\alpha \lor \beta}$ := $v_{\alpha} \lor v_{\beta}$ 
\end{lstlisting}

For each formula $\phi$ of type $\ltl{X^i\psi}$, where $\psi$ is a full-past formula, we introduce a new error bit $error_\phi$. Its monitor is defined as follows:
\begin{lstlisting}{Name}
  next($error_{\ltl{X^i \psi}}$) := case
      $error_{\ltl{X^i \psi}}$ : TRUE;
      $counter=i \land \lnot v_\psi$ : TRUE;
      TRUE : FALSE;
    esac
\end{lstlisting}

If $\phi \coloneqq \ltl{X^i G\psi}$, where $\psi$ is a full-past formula, we introduce a new error bit $error_\phi$, and we define its monitor as follows:
\begin{lstlisting}{Name}
  next($error_{\ltl{X^i G\psi}}$) := case
      $counter < i$ : FALSE;
      $\lnot error_{\ltl{X^i G\psi}} \land v_\psi$ : FALSE;
      TRUE : TRUE;
    esac
\end{lstlisting}
The same for $\phi \coloneqq \ltl{X^i(\psi_1 R \psi_2)}$:
\begin{lstlisting}{Name}
  next($error_{\ltl{X^i(\psi_1 R \psi_2)}}$) := case
      $counter < i$ : FALSE; 
      $\lnot error_{\ltl{X^i(\psi_1 R \psi_2)}} \land v^i_{\psi_1^p}$ : FALSE;
      $\lnot error_{\ltl{X^i(\psi_1 R \psi_2)}} \land v_{\psi_1} \land v_{\psi_2}$ : FALSE;
      $\lnot error_{\ltl{X^i(\psi_1 R \psi_2)}} \land v_{\psi_2}$ : FALSE;
      TRUE : TRUE;
    esac

  next($v^i_{\psi_1^p}$) := case
      $counter < i$ : FALSE; 
      $v_{\psi_1^p}$ : TRUE;
      $v^i_{\psi_1^p}$ : TRUE;
      TRUE : FALSE;
    esac
\end{lstlisting}

\begin{figure}[!ht]
\centering
\begin{adjustbox}{minipage=\linewidth,scale=.8}
  \begin{tikzpicture}[
    formula/.style={color=black},
    ph_arrow/.style={color=black, thick, ->},
  ]
    \node[formula](step1) at (4.0,7.5) {$\ltl{G(u_1 \to XXc_1) 
      \qquad\land\qquad 
      G(u_2 \to Xc_2)}$};
    \draw[black,thick] (1.45,7.0) -- (3.0,7.0);
    \draw[black,thick] (1.45,7.0) -- (1.45,7.2);
    \draw[black,thick] (3.0,7.0) -- (3.0,7.2);
    \draw[black,thick] (5.45,7.0) -- (6.85,7.0);
    \draw[black,thick] (5.45,7.0) -- (5.45,7.2);
    \draw[black,thick] (6.85,7.0) -- (6.85,7.2);
    \draw[ph_arrow,->] (2.0,6.9) -- (2.0,6.1);
    \draw[ph_arrow,->] (6.0,6.9) -- (6.0,6.1);
    \node[formula](step2) at (4.0,5.5) {$\ltl{GXX(YYu_1 \to c_1) 
      \qquad\land\qquad 
      GX(Yu_2 \to c_2)}$};
    \draw[black,thick] (0.5,5.0) -- (3.5,5.0);
    \draw[black,thick] (0.5,5.0) -- (0.5,5.2);
    \draw[black,thick] (3.5,5.0) -- (3.5,5.2);
    \draw[black,thick] (5.0,5.0) -- (7.5,5.0);
    \draw[black,thick] (5.0,5.0) -- (5.0,5.2);
    \draw[black,thick] (7.5,5.0) -- (7.5,5.2);
    \draw[ph_arrow,->] (2.0,4.9) -- (2.0,4.1);
    \draw[ph_arrow,->] (6.0,4.9) -- (6.0,4.1);
    \node[formula](step3) at (4.0,3.5) {$\ltl{XXG(YYu_1 \to c_1) 
      \qquad\land\qquad 
      XG(Yu_2 \to c_2)}$};
    \draw[black,thick] (0.5,3.0) -- (7.5,3.0);
    \draw[black,thick] (0.5,3.0) -- (0.5,3.2);
    \draw[black,thick] (7.5,3.0) -- (7.5,3.2);
    \draw[ph_arrow,->] (4.0,2.9) -- (4.0,2.1);
    \node[formula](step4) at (1.0,1.8) {\texttt{ASSIGN}}; 
    \node[formula](step5) at (1.8,1.4) {\texttt{init}($error_1$) $\coloneqq
      \bot$}; 
    \node[formula](step6) at (1.9,1.1) {\texttt{next}($error_1$) $\coloneqq
      \dots$};
    \node[formula](step4) at (6.0,1.8) {\texttt{ASSIGN}}; 
    \node[formula](step5) at (6.8,1.4) {\texttt{init}($error_2$) $\coloneqq
      \bot$}; 
    \node[formula](step6) at (6.9,1.1) {\texttt{next}($error_2$) $\coloneqq
      \dots$};
    \node[formula](step4) at (4.0,0.7) {\texttt{INVARSPEC}}; 
    \node[formula](step5) at (4.8,0.3) {$\lnot error_1 \land \lnot error_2$};
    \node[formula](step4) at (8.5,6.5) {\textsf{pastify}}; 
    \node[formula](step4) at (8.65,4.5) {\textsf{canonize}}; 
    \node[formula](step4) at (8.55,2.5) {\textsf{to SSA}}; 
  \end{tikzpicture}
  \caption{The execution of the sequence of steps: a simple example.}
  \label{fig:examplecomplete}
\end{adjustbox}
\end{figure}
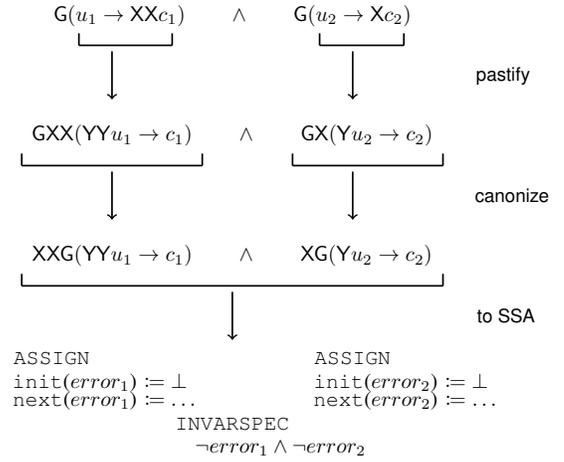

In \cref{fig:examplecomplete}, we describe the execution of all the steps described so far on a simple formula.

\begin{proposition}
\label{prop:dssasize}
  Let  $\phi$ be a canonical \PLTLEBR formula, with $|\phi| = n$. Then, there exists a deterministic SSA of size $\mathcal{O}(n)$ that accepts the same language.
\end{proposition}

\begin{theorem}
\label{th:dssasize}
  Let  $\phi$ be an \LTLEBR formula, with $|\phi| = n$, and let $M$ be the greatest constant in $\phi$. Then, there exists a deterministic SSA of size $\mathcal{O}(n^3 \cdot M^{\log_2 n + 1})$ that accepts the same language.
\end{theorem}

\begin{corollary}
\label{corol:dssasize}
  Let $\phi$ be an \LTLEBR formula with no constants, with $|\phi| = n$. Then, there exists a deterministic SSA of size $\mathcal{O}(n^3)$ that accepts the same language.
\end{corollary}
\ifarxiv
  Proofs of the above statements can be found in the appendix.
\fi


\section{%
  Solving the game on the\texorpdfstring{\\}{ }symbolic deterministic automaton%
}
\label{sec:game}
Once we have obtained the deterministic SSA $\autom(\phi)$ for an \LTLEBR formula $\phi$ with the steps described in the previous sections, we can use $\autom(\phi)$ as the arena of a two-player game between Controller and Environment in order to solve the realizability (and synthesis) problem for
$\phi$. 

Let us focus on the \emph{safety game} $G=\langle \autom(\phi), \Cset, \Uset \rangle$ (recall \cref{def:safetygame}).  Safety games have been
extensively studied, as their reachability objective makes the problem simpler than considering $\omega$-regular objectives, such as, for instance, B\"uchi and Rabin conditions.

The aim of  Controller is to choose an infinite sequence of
\emph{controllable} variables in such a way that, no matter what values for the
\emph{uncontrollable} variables are chosen by Environment, \emph{the} trace
induced by the play in $\autom(\phi)$ is \emph{safe}, that 
is, it visits only
states $s$ such that $s \models S(X)$ (see \cref{def:safetygame}). Since in our
case $\autom(\phi)$ recognizes exactly the language of $\phi$, the play
satisfies $\phi$, and thus Controller has a winning strategy for $\phi$.

Since the organization of the SYNTCOMP \cite{jacobs2017first}, many optimized
tools have been proposed in the literature to solve safety games. For this
reason, we chose to use a safety synthesizer as a black box. The majority of
these tools accept as input a symbolic arena described in terms of and-inverter
graphs (or AIGER format \cite{biere2011aiger}), so we provide a simple utility
to obtain the AIGER representation of \emph{functional} SMV modules, that is, SMV
modules with the transition relation expressed only in terms of \textsf{ASSIGN}
statements, such as the ones resulting from our encoding. The AIGER model is then
given as input to the chosen safety synthesizer, completing the process outlined in 
\cref{fig:main}.

The next theorem states the complexity of the procedure.
\begin{theorem}
  The realizability problem for \LTLEBR belongs to \EXPTIME[2]. If no constant is admitted, it belongs to \EXPTIME.
\end{theorem}
\begin{IEEEproof}
  We first show that the proposed algorithm, as described in \cref{fig:main}, belongs to
  \EXPTIME[2] for generic \LTLEBR formulas. It is easy to see that the time complexity of all the steps matches their space complexity. Therefore, we have an algorithm to turn an \LTLEBR formula $\phi$ into an equivalent
  deterministic SSA $\autom(\phi)$ whose time complexity
  is $\mathcal{O}(n^3 \cdot M^{\log_2 n + 1})$, where $n = |\phi|$ and $M$ is
  the greatest constant in $\phi$. Since $\autom(\phi)$ is symbolically
  represented, it can be turned into an explicit automaton
  $\autom^\prime(\phi)$ of size at most exponential in the size of
  $\autom(\phi)$, that is, $|\autom^\prime(\phi)| \in \mathcal{O}(2^{n^3 \cdot
  M^{\log_2 n + 1}})$. Finally, the time complexity of reachability games is \emph{linear} 
  in the size of the arena \cite{de2007concurrent}, and thus
  the overall time complexity of the realizability problem for \LTLEBR is
  \EXPTIME[2]. If no constant is admitted, then, by \cref{corol:dssasize},
  $|\autom^\prime(\phi)| \in \mathcal{O}(2^{n^3})$, and the complexity becomes \EXPTIME.
\end{IEEEproof}

\subsection*{Comparison with Safety \LTL}
\label{sub:ssyftdelta}
It is interesting to briefly compare the proposed procedure for realizability
to the one used by the \ssyft tool for Safety \LTL specifications
\cite{zhu2017symbolic}. In that tool, the negation of the initial formula  is
first translated into first-order logic over finite words and then transformed
into deterministic automata using the tool \mona \cite{henriksen1995mona},
which uses the classical subset construction to determinize automata over
finite words.  Finally, \ssyft uses the classical backward fixpoint iteration
to compute the set of winning states over the \DFA.  It is worth to notice that
the way \mona represents automata is \emph{not} fully symbolic: the set of
states is explicitly represented, while it uses a BDD for each pair of states
in order to represent symbolically the transitions between the two corresponding
states.  In contrast of subset construction, our solution performs the
pastification of full-bounded formulas. Most importantly, our construction of
deterministic monitors is carried out in a fully symbolic way.


\section{Experimental Evaluation}
\label{sec:expeval}
We implemented the proposed procedure (see \cref{fig:main}) in
a tool called
\tool.\footnote{\url{http://users.dimi.uniud.it/~luca.geatti/tools/ebrltlsynth.html}}
The transformation from \LTLEBR to deterministic SSA together with the translation
to AIGER has been implemented inside the \nuxmv model checker
\cite{cavada2014nuxmv}. As the backend for solving the safety game, we have
chosen the SAT-based tool \demiurge \cite{bloem2014sat}.

We tested our tool on a set of scalable benchmarks divided in four categories
(the propositional atoms starting with the letter \emph{c} are controllable,
while those starting with the letter \emph{u} are uncontrollable):
\begin{enumerate}
  \item the first category is generated by the realizable formula:
    \begin{align}
      \ltl{G(c_0 && XG(c_1 && \dots && X^nG(c_n && u) \dots ))}
    \end{align}
  \item the second category is generated by the realizable formula: 
    \begin{align}
      \ltl{G((c_0 || u_0) && XG((c_1 || u_1) && \dots && X^nG((c_n ||
      u_n))\dots))}
    \end{align}
  \item the third category is generated by the unrealizable formula:
    \begin{align}
      \ltl{G(c) && \bigvee_{i=1}^{n}G(\bigwedge_{j=0}^{i}u_i)}
    \end{align}
  \item the fourth category is generated by the unrealizable formula:
    \begin{align}
      \ltl{c && \bigwedge_{i=1}^{n}X^i(u_i || u_{i+1})}
    \end{align}
\end{enumerate}
Each category contains the respective scalable formula for $n \in [1,200]$, for
a total of $800$ benchmarks, half of which is realizable and the other half is
unrealizable. We set a timeout of 180 seconds for each benchmark.  We compared
\tool with \ltlsynt \cite{jacobs5th}, \strix \cite{meyer2018strix} and \ssyft
\cite{zhu2017symbolic}. The first two tools solve the realizability and
synthesis problems for full \LTL and are based on a translation to parity
games. \ltlsynt uses SPOT \cite{duret2016spot} for efficient translation and
manipulation of automata.  \strix implements several optimizations like
specification splitting, that enables to split the initial formula in safety,
co-safety, B\"uchi, and co-B\"uchi subformulas and speeds up the process of
solving of the game.  On the contrary, \ssyft solves the realizability problem
for specifications written in Safety \LTL (see \cref{sub:ssyftdelta} for a brief
description of the \ssyft tool).

For realizability, we tested all the tools in their sequential configurations.
\ltlsynt has two sequential configurations, which differ on whether the split
of actions into Controller's and Environment's ones is performed before or
after the determinization. \strix has two sequential modes as well, depending
on the kind of search on the arena (depth-first for the first configuration and
with a priority queue for the second).  \ssyft and \tool have only one
configuration.

\begin{figure}[t]
  \centering
  \includegraphics[width=0.8\linewidth]{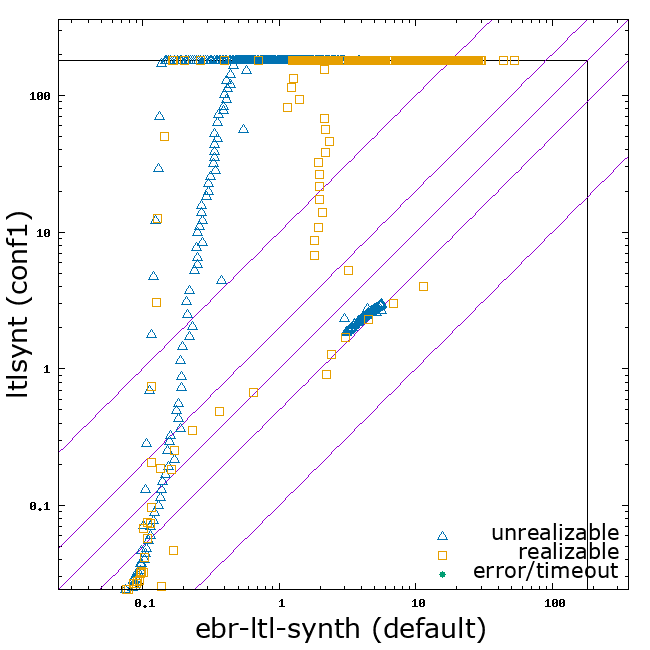}
  \caption{\tool vs \ltlsynt (first conf.) on all scalable benchmarks.}
  \label{fig:ebrltlsynt1}
\end{figure}

\begin{figure}[t]
  \centering
  \includegraphics[width=0.8\linewidth]{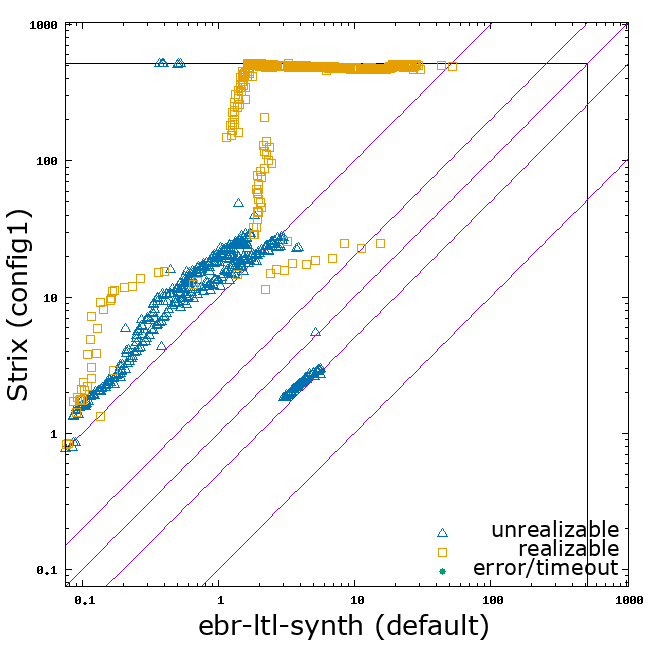}
  \caption{\tool vs \strix on all scalable benchmarks.}
  \label{fig:ebrstrix1}
\end{figure}

\begin{figure}[t]
  \centering
  \includegraphics[width=0.8\linewidth]{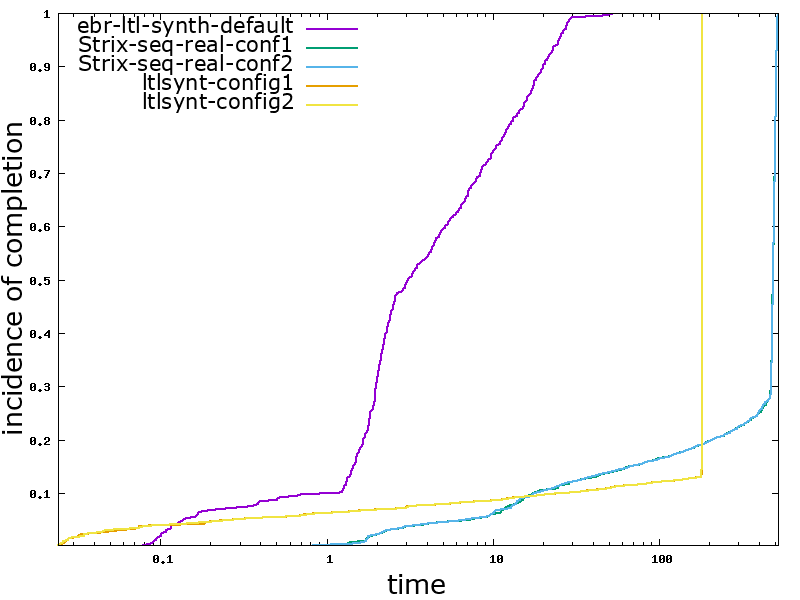}
  \caption{Survival plot for realizable scalable benchmarks.}
  \label{fig:realcdf}
\end{figure}

\begin{figure}[t]
  \centering
  \includegraphics[width=0.8\linewidth]{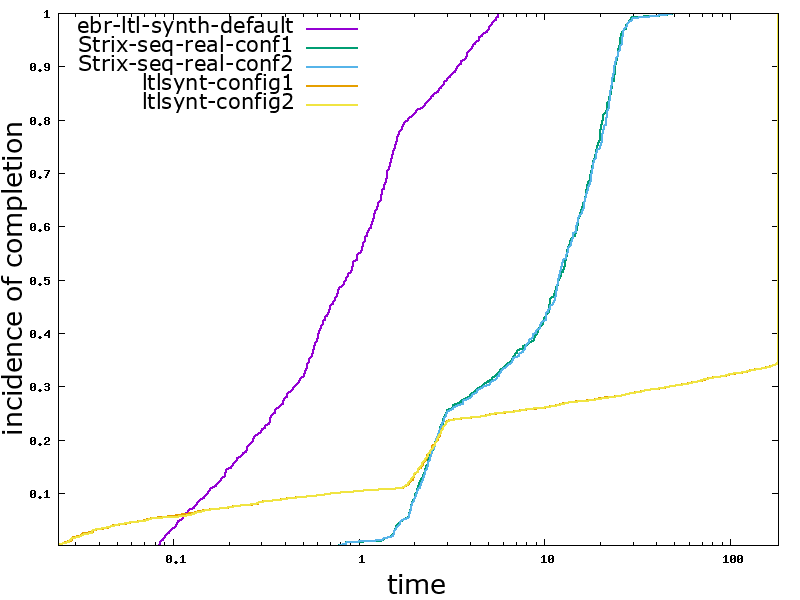}
  \caption{Survival plot for unrealizable scalable benchmarks.}
  \label{fig:unrealcdf}
\end{figure}

\Cref{fig:ebrltlsynt1} shows the outcomes of the  comparison between \tool and
the best configuration of \ltlsynt: it can be clearly seen that, for both
realizable and unrealizable formulas, \ltlsynt presents an exponential blow-up
in the solving time that is avoided by \tool. \Cref{fig:ebrstrix1} compares
\tool with the best configuration of \strix: while for realizable formulas
there is an exponential blow up of \strix avoided by \tool, it is interesting
to note that for the unrealizable benchmarks the difference between the solving
time of the two tools is linear, mostly showing a 10x improvement in favor of
\tool.  The survival plots for the set of realizable and unrealizable scalable
benchmarks are shown in \cref{fig:realcdf,fig:unrealcdf}, respectively. 

\begin{figure}[t]
  \centering
  \includegraphics[width=0.8\linewidth]{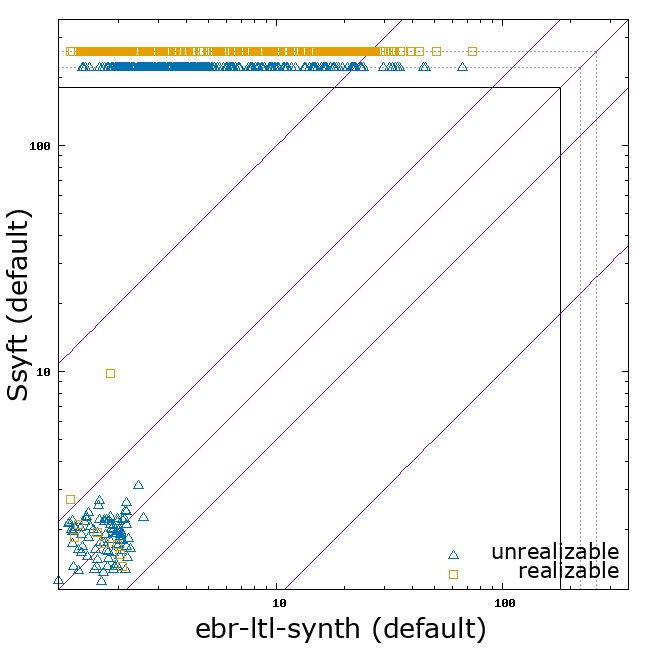}
  \caption{\tool vs \ssyft on scalable benchmarks.}
  \label{fig:ebrssyftscalable}
\end{figure}%

\begin{figure}[t]
  \centering
  \includegraphics[width=0.8\linewidth]{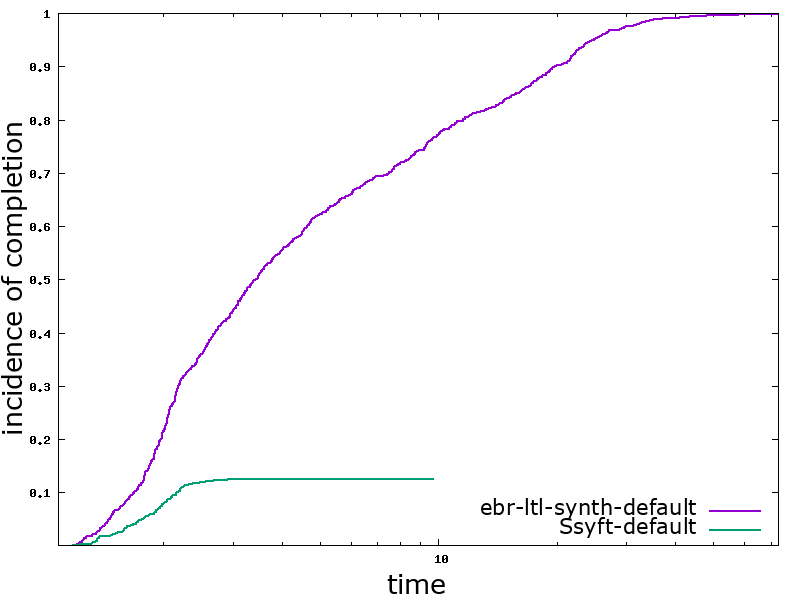}
  \caption{Survival plot for \tool and \ssyft on scalable benchmarks.}
  \label{fig:ebrssyftscalablecdf}
\end{figure}%

\begin{figure}[t]
  \centering
  \includegraphics[width=0.8\linewidth]{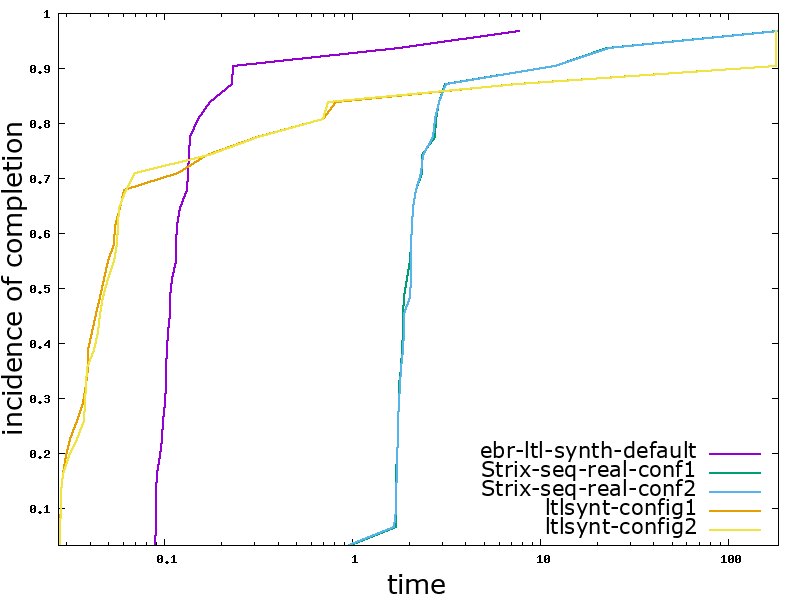}
  \caption{Survival plot for SYNTCOMP benchmarks.}
  \label{fig:syntcompcdf}
\end{figure}%

\begin{figure}[t]
  \centering
  \includegraphics[width=0.8\linewidth]{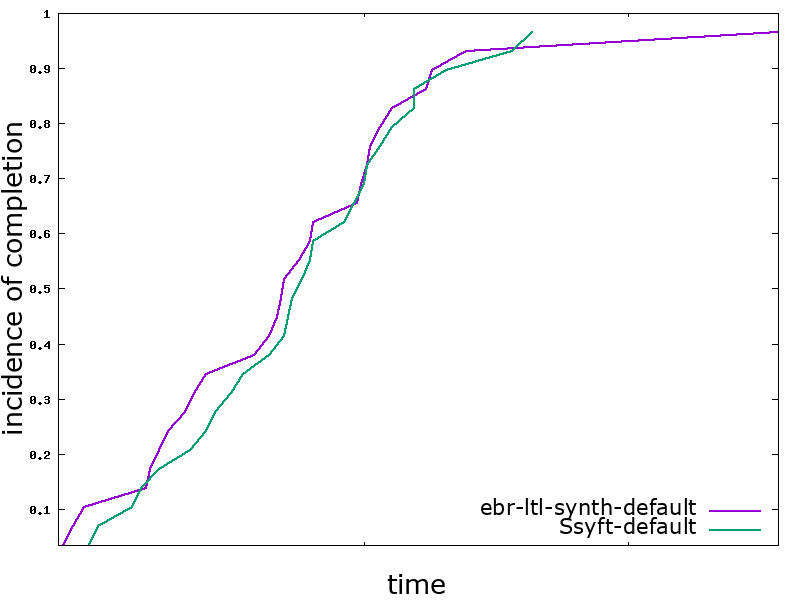}
  \caption{Survival plot for \tool and \ssyft on SYNTCOMP benchmarks.}
  \label{fig:syntcompcdfssyft}
\end{figure}%

The outcomes of the comparison between \tool and \ssyft are shown in
\cref{fig:ebrssyftscalable}. Here the three lines near the sides of the figure
correspond to \emph{timeouts} (the solid black line), \emph{memouts} for
unrealizable benchmarks and \emph{memouts} for realizable benchmarks (the
dotted lines). It can be noticed that \ssyft reaches a memory out for the vast
majority of benchmarks.  For instance, on both the realizable categories,
\ssyft reaches the first memout with $n=7$. As for the unrealizable benchmarks,
on the third category, \ssyft reaches the first memout with $n=36$, while for
the fourth category with $n=59$.  This is due to \mona, which is not able to
build the (explicit) \DFA for the (negation of the) initial
specification\footnote{We point out that in some cases, like in the fourth
category for $n \ge 60$, \mona's memouts are due to its parser.}.  This is an
important hint about the use of \emph{fully symbolic} techniques for the
representation of automata, like the one of \tool , as in many cases they can
avoid an exponential blowup of the automata' state space.  The survival plot
between \tool and \ssyft is shown in
\cref{fig:ebrssyftscalablecdf}\footnote{The reason why we do not have a single
survival plot comparing all the four tools is that \ssyft could not have been
compiled for the same platform as the others, due to issues with its source
code.}.
\ifarxiv
  The rest of the plots for realizability of scalable benchmarks can be found 
  in the appendix.
\else
  The rest of the plots for realizability of scalable benchmarks can be found 
  in \cite{CimattiGGMT20arxiv}.
\fi

In addition to these scalable formulas, from the benchmarks of SYNTCOMP
\cite{jacobs2017first}, we filtered the formulas that belong to \LTLEBR: this
resulted into a set of 29 formulas. The survival plot showing the comparison
with \ltlsynt and \strix is shown in \cref{fig:syntcompcdf}, while the
comparison with \ssyft is shown in \cref{fig:syntcompcdfssyft}.  It is
interesting to see that, on the SYNTCOMP benchmarks, the results of \tool and
\ssyft are comparable.

As for the synthesis problem, once a specification is found to be realizable,
all the three tools produce a strategy as a witness: this strategy is in the
form of an and-inverter graph whose  input bits are only the starting
uncontrollable variables. Often, a strategy of this kind can be minimized by
using logic synthesis tools (like \abc \cite{brayton2010abc}) as black-box. In
the particular case of the tools considered in this section, they all use a
separate logic synthesizer as black box, with different configurations to
minimize the strategy. Therefore, we do not compare the size of the strategies
found by the three tools, since such a comparison would add nothing about the
methods implemented by the tools but would rather compare their backends.


\section{Conclusions}
\label{sec:end}
In this paper, we introduce the logic \LTLEBR, a fragment of \LTL
that combines formulas with only bounded operators and a particular combination
of universal unbounded temporal operators. We focus on the realizability and
reactive synthesis problems for this logic. 
The main contribution is
a \emph{fully symbolic} translation from any \LTLEBR formula to
a \emph{deterministic} symbolic safety automaton on infinite words. The process
applies a pastification step and a set of rules to reach a canonical
form for \LTLEBR formulas. The realizability is then decided by solving
a safety game on the arena represented by the automaton.
We first showed that realizability for \LTLEBR belongs to \EXPTIME[2],
but drops to \EXPTIME if no constant is used. Then, we
implemented the proposed procedure in a tool, whose experimental evaluation
revealed very good performance against tools for realizability and
synthesis of full \LTL and Safety \LTL specifications.

\smallskip

As a future development of this line of work, we believe that the translation
from \LTLEBR to deterministic SSA may provide many benefits in the context of
\emph{symbolic model checking} as well, since the search of the state space
could benefit from a deterministic representation of the automaton for the
formula~\cite{SebastianiT03}.
On the automata construction side, an interesting development would be to keep
the symbolic bounds during pastification and monitor construction, without, for
instance, expanding $\ltl{X^i\alpha}$ into $i$ nested \emph{next} operators.
On the expressiveness side, we want to study in which ways \emph{assumptions}
can be integrated into \LTLEBR.
Last but not least, we aim at checking whether the synthesis problem for more
expressive logics, like, for instance, \LTL, can be reduced to the synthesis
problem for \LTLEBR, for example checking whether it is possible to use
\LTLEBR for solving the safety problems originated from \emph{bounded
synthesis} techniques.

\paragraph*{Acknowledgments}
The authors want to thank all the anonymous reviewers of FMCAD 2020 for the
insightful comments on a preliminary version of this paper.

\ifarxiv
\else
  \newpage
\fi

\bibliographystyle{splncs04}
\bibliography{biblio}

\clearpage
\appendices
\ifarxiv

\section{Proofs}

%
%

\begin{proposition}[Soundness of pastification]
  \label{app:prop:pastification}
  Let $\varphi$ be a \LTLFB formula. For all state sequences
  $\sigma \in (2^\Sigma)^\omega$, all $i \in \N$, and all $d\ge D(\phi)$, 
  it holds that:
  \begin{align}
    \sigma,i \models \varphi \ &\Leftrightarrow \ 
    \sigma,i \models
    \ltl{X^d \Pi(\varphi,d)}
  \end{align}
\end{proposition}
\begin{IEEEproof}
\label{app:proof:pastification}
  The proof goes by structural induction over $\varphi$. As the base case, 
  consider a proposition $p\in\Sigma$, and since $D(p)=0$, consider any 
  $d\ge0$. It holds that $\sigma,i\models p$ if and only if 
  $\sigma,i\models \ltl{X^d Y^d p}$, which is equivalent to say that
  $\sigma,i+d \models \ltl{Y^d p}$, hence $\sigma,i+d\models \Pi(p,d)$. For the 
  inductive case, we consider multiple cases:
  \begin{enumerate}
  \item if $\phi\equiv\ltl{X\phi_1}$, consider any $d\ge D(\ltl{X\phi_1})$. By
        the semantics of the \emph{tomorrow} operator,
        $\sigma,i\models\ltl{X\phi_1}$ is equivalent to
        $\sigma,i+1\models\phi_1$, which, by the inductive hypothesis, is
        equivalent to $\sigma,i+1+t\models \Pi(\phi_1, t)$ for all $t\ge
        D(\phi_1)$. Since $D(\ltl{X\phi_1})=D(\phi_1)+1$, the above is
        equivalent to $\sigma,i+d\models \Pi(\phi_1, d-1)$, hence
        $\sigma,i+d\models\Pi(\ltl{X\phi_1},d)$, for all $d\ge
        D(\ltl{X\phi_1})$.
    \item if $\phi\equiv \ltl{\phi_1 U^{[a,b]} \phi_2}$, consider any 
        $d\ge D(\phi)$. The following equivalences hold:
        \begingroup
          \def\why#1{\tag*{\small \emph{#1}}}
          \begin{align}
            &\sigma,i\models\ltl{\phi_1 U^{[a,b]} \phi_2}  \\
          \iff\; & 
            \exists j \in [a,b] \bigl( \sigma,i+j\models \phi_2 \land \\ 
              &\qquad\forall w\in[0,j) \suchdot \sigma,i+w\models\phi_1 \bigr) \\
            \why{semantics of \emph{until}} \\
          \iff\; &
            \exists j\in[a,b](
                \sigma,i+j+d-b\models\Pi(\phi_2,d-b) \land\, \\
              &\qquad\forall w\in[0,j) \suchdot
              \sigma,i+w+d-b\models\Pi(\phi_1,d-b) 
            )
            \why{by the inductive hypothesis,}\\[-0.5ex]
            \why{since $D(\phi)\ge D(\phi_1)$ and $D(\phi)\ge D(\phi_2)$} \\
          \iff\; &
            \exists t\in[0,b-a](
                \sigma,i-t+d\models\Pi(\phi_2,d-b) \land\, \\
              &\quad\forall w'\in [0,b-t-1] \suchdot \\
              &\quad\sigma,i-t-w'+d-1 \models\Pi(\phi_1,d-b)
            )
            \why{since $w'=b-t-w-1$ and $t=b-j$} \\
          \iff\; &
            \exists t\in[0,b-a](
                \sigma,i+d\models\ltl{Y^t\Pi(\phi_2,d-b)} \land\, \\ 
                &\qquad\sigma,i+d\models\ltl{Y^tH^{\le b-t-1}Y\Pi(\phi_1,d-b)}
            )
            \why{semantics of \emph{yesterday} and \emph{historically}}\\
          \iff\; & \sigma,i+d \models \\ 
            &\qquad\bigvee_{t=0}^{b-a} \ltl{
              Y^t\bigl(
                \Pi(\phi_2,d-b) \land H^{\le b-t-1}Y\Pi(\phi_1,d-b)
              \bigr)
            }
              \why{conjunction and disjunction}\\
            \iff\; & \sigma,i+d\models\Pi(\ltl{\phi_1 U^{[a,b]}\phi_2},d)
          \end{align}
        \endgroup
  \end{enumerate}
  This concludes the proof. \IEEEQEDhere
\end{IEEEproof}

\newpage
\begin{proposition}
\label{app:prop:pastsize}
  Let $\phi$ be a \LTLFB formula. Then, $\pastify(\phi)$ is a formula of size
  $\mathcal{O}(n^2 \cdot M^{\log_2 n + 1})$, where $n = |\phi|$ and $M$ is
  the greatest constant in $\phi$.
\end{proposition}
\begin{IEEEproof}
\label{app:proof:pastsize}
  We first give a bound for the $\Pi(\cdot)$ operator. It holds that:
  \begin{itemize}
    \item $|\Pi(p,d)| = \mathcal{O}(p)$ for each $p \in \Sigma$;
    \item $|\Pi(\lnot\phi,d)| = |\Pi(\phi,d)| + 1$;
    \item $|\Pi(\phi_1 \land \phi_2,d)| = |\Pi(\phi_1,d)| + |\Pi(\phi_2,d)|
      + 1$;
    \item $|\Pi(\ltl{X\phi_1},d)| \le |\Pi(\phi_1,d)| + 1$;
  \end{itemize}
  and 
  \begin{align}
    |\Pi(\ltl{\phi_1 U^{[a,b]} \phi_2},d)| 
    &\le 1+ \sum\limits_{i=0}^{M}(i + |\Pi(\phi_2,d-i)| + \\
      &\qquad(M-i) + |\Pi(\phi_1,d-i)|) \\
    &\le 1+ \sum\limits_{i=0}^{M}(M + |\Pi(\phi_2,d-i)| + \\
     &\qquad|\Pi(\phi_1,d-i)|) \\
    &\le 1+ M^2 + M|\Pi(\phi_2,d)| + M|\Pi(\phi_1,d)|
  \end{align}
  Since the case for the bounded until operator dominates all the others, we
  have that $|\Pi(\phi,d)| \le 1+ M^2 + M|\Pi(\phi_2,d)| + M|\Pi(\phi_1,d)|$,
  where $|\phi| = 1+|\phi_1|+|\phi_2|$. Without loss of generality, we can
  assume that $|\phi_1| = |\phi_2| = \frac{|\phi|-1}{2}$; in this way, the
  recurrence equation $S(n)$ describing the space required for $|\Pi(\phi,d)|$,
  with $n=|\phi|$, is the following:
  \begin{equation}
    S(n)=
    \begin{cases}
      \mathcal{O}(d) & \text{if}\ n=1 \\
      2M\cdot S(\frac{n}{2}) + \mathcal{O}(M^2) & \text{otherwise}
    \end{cases}
  \end{equation}
  By unrolling the equation for $i$ steps, we have that $S(n) = (2M)^i \cdot
  S(\frac{n}{2^i}) + \mathcal{O}(M^i)$. For $i = \log_2n$, the equation amounts
  to:
  \begin{align}
    S(n) &= (2M)^{\log_2n} \cdot S(1) + \mathcal{O}(M^{\log_2n}) \\
    &= d\cdot(2M)^{\log_2n} + \mathcal{O}(M^{\log_2n})
  \end{align}
  Since $\pastify(\phi)$ is defined as $X^{d}\Pi(\phi,d))$ where $d = D(\phi)$,
  it holds that:
  \begin{align}
    \pastify(\phi) 
    &\le d + d \cdot (2M)^{\log_2n} + \mathcal{O}(M^{\log_2n}) \\
    &\le Mn + Mn \cdot (2M)^{\log_2n} + \mathcal{O}(M^{\log_2n}) \\
    &\qquad \text{ since } d \le Mn \\
    &\in \mathcal{O}(M\cdot n\cdot (2M)^{\log_2n}) \\
    &\in \mathcal{O}(M\cdot n\cdot 2^{\log_2 n} \cdot M^{\log_2 n}) \\
    &\in \mathcal{O}(n^2 \cdot M^{\log_2 n + 1}) \\
  \end{align}
\end{IEEEproof}

\newpage
\begin{lemma}[Strong equivalence for the rules]
\label{app:lemma:equivalences}
  Let $\psi$, $\psi_1$, $\psi_2$ and $\psi_3$ be \LTLFP formulas. For all state
  sequences $\sigma$ and for all positions $i \in \N$, it holds that:
  \begin{itemize}[leftmargin=30pt]
    \item[$R_1$:] 
      $\ltl{\sigma,i\models X(\psi_1 \land \psi_2) \iff 
        \sigma,i \models X\psi_1 \land X\psi_2 }$
    \item[$R_2$:]
      $\ltl{\sigma,i \models \psi R (\psi_1 \land \psi_2) \iff 
        \sigma, i \models \psi R \psi_1 \land \psi R \psi_2 }$
    \item[$R_3$:]
      $\ltl{\sigma,i \models (X^i \psi_1) R (X^j \psi_2)} \iff$
      \begin{equation}
        \sigma,i \models
      \begin{cases}
        \ltl{X^i(\psi_1 R (Y^{i-j}\psi_2))}
          & \mbox{ if } i>j \\ 
        \ltl{X^j((Y^{j-i}\psi_1) R \psi_2)} 
          & \mbox{ otherwise}
      \end{cases}
      \end{equation}
    \item[$R_4$:]
      $\sigma,i \models \ltl{(X^i \psi_1)R(X^j(\psi_2 R \psi_3))} \Leftrightarrow$
      \begin{equation}
        \sigma,i \models 
      \begin{cases}
        \ltl{X^i( \psi_1 R ((Y^{i-j}\psi_2) R (Y^{i-j} \psi_3) ) )}
          \\ \quad \mbox{ if } i>j \\
        \ltl{X^j( (Y^{j-i}(\psi_1 \land \true)) R (\psi_2 R \psi_3) )} 
          \\ \quad \mbox{ otherwise }
      \end{cases}
      \end{equation}
    \item[$R_5$:]
      $\ltl{\sigma,i \models GX^iG\psi \iff 
        \sigma, i \models X^iG\psi}$
    \item[$R_6$:]
      $\ltl{\sigma,i \models  GX^i(\psi_1 R \psi_2)\iff 
        \sigma, i \models X^i G \psi_2}$
    \item[$R_7$:]
      $\ltl{(X^i\psi_1) R (X^j G \psi_2 )} \iff$
      \begin{equation}
        \sigma,i \models
      \begin{cases}
        \ltl{X^i G Y^{i-j} \psi_2}
          & \mbox{ if } i>j \\ 
        \ltl{X^j G \psi_2} 
          & \mbox{ otherwise}
      \end{cases}
      \end{equation}
    \item[$R_{flat}$:]
      $\ltl{\sigma,0 \models X^i(\psi_1 R (\psi_2 R (\dots (\psi_{n-1}
        R \psi_n)\dots)))} 
        \Leftrightarrow 
        \ltl{\sigma,0 \models
        X^i( (\psi_{n-1} \land O(\psi_{n-2} \land \dots O(\psi_1
        \land Y^i \true)\dots))) R \psi_n )} \ \forall n \ge 3$
  \end{itemize}
\end{lemma}
\begin{IEEEproof}
  Before starting the proof, we remark that the claim of this lemma not only
  asks for proving the \emph{equivalence} between the left- and the right-hand
  side of the rules, but requires to prove the \emph{strong equivalence}
  between the two, \ie that for all the state sequences $\sigma$ and for all the
  positions $i$, $\sigma$ is a model starting from position $i$ of the
  left-hand formula iff $\sigma$ is a model starting from position $i$ of the
  right-hand formula. Equivalence is a special case of strong equivalence by
  considering only $i=0$. In our case, the \emph{necessity} of considering
  strong equivalence is due to the fact that the left-hand side of the rules
  (except for $R_{flat}$, for which we require only the equivalence) can appear
  as subformulas of the original $\phi$ on which we apply the \canonize
  algorithm, and thus it can be interpreted potentially on any position $i$.
  Since we want to maintain the equivalence between $\phi$ and
  $\canonize(\phi)$, we have to make sure that each subformulas is strongly
  equivalent to the one by which it is replaced during the applications of the
  rules. The only exception is the $R_{flat}$ rule, which is applied only to
  top-level conjuncts or disjuncts, and thus we can require for it to maintain
  only the equivalence.
  
  Initially we prove the first two points (\ie $R_1$ and $R_2$). For the
  $R_1$ rule, the following steps hold:
  \begin{align}
    &\sigma,i \models \ltl{X(\psi_1 \land \psi_2)} \\
    \Leftrightarrow \ &\sigma,i+1 \models \psi_1 \land \psi_2 \\
    \Leftrightarrow \ &\sigma,i+1 \models \psi_1 \land \sigma,i+1 \models \psi_2 \\
    \Leftrightarrow \ &\sigma,i \models \ltl{X \psi_1} \land \sigma,i \models \ltl{X \psi_2} \\
    \Leftrightarrow \ &\sigma,i \models \ltl{X\psi_1 \land X\psi_2}
  \end{align}
  Consider rule $R_2$. We first prove that $\sigma, s \models \ltl{\psi
  R (\phi_1 \land \phi_2)}$ implies $\sigma,s \models \ltl{\psi R \phi_1 \land
  \psi R \phi_2}$, for all state sequences $\sigma$ and for all positions $s$.
  Let $\sigma$ be a state sequence and let $s \in \N$ be a position such that
  $\sigma,s \models \ltl{\psi R (\phi_1 \land \phi_2)}$. We divide in cases:
  \begin{enumerate}
    \item if $\forall i\ge s. (\sigma,i \models \phi_1 \land \phi_2)$, then
      $\forall i\ge s. \sigma,i \models \phi_1$ and $\forall i\ge s. \sigma,i
      \models \phi_2$. Thus, $\sigma,s \models \ltl{\psi R \phi_1}$ and
      $\sigma,s \models \ltl{\psi R \phi_2}$, that is $\sigma,s \models
      \ltl{\psi R \phi_1 \land \psi R \phi_2}$.
    \item if $\exists i \ge s. (\sigma,i \models \psi \land
      \forall s \le j \le i. \sigma,j \models (\phi_1 \land \phi_2))$
      \begin{align}
        &\Leftrightarrow \exists i \ge s. (\sigma,i \models \psi \land
          \forall s \le j \le i. (\sigma,j \models \phi_1) \land \\
          &\qquad\qquad \forall s \le k \le i. (\sigma,k \models \phi_2)) \\
        &\Rightarrow \exists i \ge s. (\sigma,i \models \psi \land 
          \forall s \le j \le i. (\sigma,j \models \phi_1)) \land \\
          &\qquad \exists i \ge s. (\sigma,i \models \psi \land \forall 0 \le j \le i.
          (\sigma,j \models \phi_2)) \\
        &\Leftrightarrow \sigma,s \models \ltl{\psi R \phi_1 \land \psi R \phi_2}
      \end{align}
  \end{enumerate}
  We now prove the opposite direction, that is $\sigma,s \models \ltl{\psi
  R \phi_1 \land \psi R \phi_2}$ implies $\sigma,s \models \ltl{\psi R (\phi_1
  \land \phi_2)}$, for all state sequences $\sigma$ and for all positions $s$.
  Let $\sigma$ be a state sequence and let $s \in \N$ such that $\sigma,s
  \models \ltl{\psi R \phi_1 \land \psi R \phi_2}$. We divide again in cases:
  \begin{enumerate}
    \item if $\forall i \ge s. (\sigma,i \models \phi_1) \land \forall i\ge s.
      (\sigma,j \models \phi_2)$, then $\forall i \ge s. (\sigma,i \models
      \phi_1 \land \phi_2)$ and thus $\sigma,s \models \ltl{\psi R (\phi_1
      \land \phi_2)}$.
    \item if $\forall i \ge s. (\sigma,i \models \phi_1)$ and $\exists i \ge s.
      (\sigma,i \models \psi \land \forall s \le j \le i. \sigma,j \models
      \phi_2)$, then $\exists i \ge s (\sigma,i \models \psi \land \forall
      s \le j \le i . \sigma,j \models (\phi_1 \land \phi_2))$, that is
      $\sigma,s \models \ltl{\psi R (\phi_1 \land \phi_2)}$.
    \item if $\exists i \ge s. (\sigma,i \models \psi \land \forall s \le j \le
      i . \sigma,j \models \phi_1)$ and $\forall i \ge s. (\sigma,i \models
      \phi_2)$, then $\exists i\ge s. (\sigma,i \models \psi \land \forall
      s \le j \le i. \sigma,k \models \phi_1 \land \phi_2)$, that is $\sigma,s
      \models \ltl{\psi R (\phi_1 \land \phi_2)}$.
    \item consider the case such that $\exists l \ge s .(\sigma,l \models \psi
      \land \forall s \le j \le l . \sigma,j \models \phi_1)$ and $\exists
      k \ge s. (\sigma,k \models \psi \land \forall s \le j \le k . \sigma,j
      \models \phi_2)$. Let $i = \min(l,k)$: then $\sigma,i \models \phi$ and
      $\forall s \le j \le i.(\sigma,j \models \phi_1 \land \phi_2)$, that is
      $\sigma,s \models \ltl{\psi R (\phi_1 \land \phi_2)}$.
  \end{enumerate}
  This concludes the proof for the $R_2$ rule.

  Before proving the cases of the remaining rules, we define and prove the
  following auxiliary \emph{strong equivalences}. For all state sequences
  $\sigma$ and for all positions $i$, it holds that:
  \begin{itemize}
    \item[$\bar{R_1}$:] 
      $\ltl{\sigma,i \models \psi_1 R (X^i \psi_2) \ \iff \ \sigma,i \models
      X^i((Y^i\psi_1) R \psi_2)}$
    \item[$\bar{R_2}$:]
      $\ltl{\sigma,i \models (X^i\psi_1)R\psi_2 \ \iff \ \sigma,i \models
      X^i(\psi_1 R (Y^i\psi_2))}$
    \item[$\bar{R_3}$:]
      $\ltl{\sigma,i \models Y^iX^i\psi \ \iff \ \sigma,i \models \psi \land
      Y^i\true}$
    \item[$\bar{R_4}$:]
      $\ltl{\sigma,i \models Y^i(\psi_1 R \psi_2) \ \iff \ \sigma,i \models
      (Y^i\psi_1) R (Y^i \psi_2)}$
    \item[$\bar{R_5}$:]
      $\ltl{\sigma,i \models G G \psi \ \iff \ \sigma,i \models G \psi}$
    \item[$\bar{R_6}$:]
      $\ltl{\sigma,i \models G(\psi_1 R \psi_2) \ \iff \ \sigma,i \models
      G \psi_2}$
    \item[$\bar{R_7}$:]
      $\ltl{\sigma,i \models \psi_1 R (G\psi_2) \ \iff \ \sigma,i \models
      G \psi_2}$
  \end{itemize}
  These will help proving the cases for $R_3$-$R_7$.

  Consider the case for rule $\bar{R_1}$. We first prove that $\ltl{\sigma,s
  \models \psi_1 R (X^i\psi_2)}$ implies $\ltl{\sigma,s \models X^i((Y^i\psi_1)
  R \psi_2)}$, for all state sequences $\sigma$ and all positions $s$. Let
  $\sigma$ be a state sequence and let $s \in \N$ such that $\sigma, s \models
  \ltl{\psi_1 R (X^i\psi_2)}$. We divide in cases:
  \begin{enumerate}
    \item if $\forall j \ge s. \sigma,j \models \ltl{X^i\psi_2}$, then 
      \begin{align}
        &\Leftrightarrow \forall j \ge s+i . \sigma,j \models \psi_2 \\
        &\Rightarrow \sigma,s+i \models \ltl{(Y^i\psi_1) R \psi_2} \\
        &\Leftrightarrow \sigma,s \models \ltl{X^i((Y^i \psi_1)R\psi_2)}
      \end{align}
    \item if $\exists j \ge s. (\sigma,j \models \psi_1 \land \forall s \le
      k \le j . \sigma,k \models \ltl{X^i\psi_2})$, then $\exists j \ge s.
      (\sigma,j+i \models \ltl{Y^i\psi_1} \land \forall s+i \le k \le j+i.
      \sigma,k \models \psi_2)$, which in turn means that $\sigma,s+i \models
      \ltl{(Y^i\psi_1) R \psi_2}$, that is $\sigma,s \models
      \ltl{X^i((Y^i\psi_1) R \psi_2)}$.
  \end{enumerate}
  We now prove the opposite direction, that is $\sigma,s \models
  \ltl{X^i((Y^i\psi_1) R \psi_2)}$ implies $\sigma,s \models \ltl{\psi_1
  R (X^i\psi_2)}$, for all state sequences $\sigma$ and all positions $s$. Let
  $\sigma$ be a state sequence and let $s \in \N$ such that $\sigma, s \models
  \ltl{X^i((Y^i \psi_1) R \psi_2)}$. We divide again in cases:
  \begin{enumerate}
    \item if $\forall j \ge s+i. (\sigma,j \models \ltl{\psi_2})$, then $\forall
      j \ge s. (\sigma,j \models \ltl{X^i\psi_2})$ and thus $\sigma \models
      \ltl{\psi_1 R (X^i\psi_2)}$.
    \item if $\exists j \ge s+i. (\sigma,j \models \ltl{Y^i\psi_1} \land \forall
      s+i \le k \le j. \sigma,k \models \psi_2)$, then:
      \begin{align}
        &\Leftrightarrow \exists j \ge s+i . (\sigma,j-i \models
        \ltl{X^iY^i\psi_1} \land
          \forall s \le k \le j-i. \\
          &\qquad\sigma,k \models \ltl{X^i\psi_2}) \\
        &\Leftrightarrow \exists j \ge s+i . (\sigma,j-i \models
        \ltl{\psi_1} \land \forall s \le k \le j-i.  \sigma,k \models
        \ltl{X^i\psi_2}) \\
        &\Leftrightarrow \sigma,s+i \models \ltl{Y^i(\psi_1 R (X^i\psi_2))} \\
        &\Leftrightarrow \sigma,s \models \ltl{\psi_1 R (X^i\psi_2)}
      \end{align}
  \end{enumerate}
  This concludes the proof for the rule $\bar{R_1}$. The proof for the
  $\bar{R_2}$ rule is specular.

  Consider the $\bar{R_3}$ case. We first prove that $\sigma,s \models
  \ltl{Y^iX^i\psi}$ implies $\sigma,s \models \ltl{\psi \land Y^i\true}$, for
  all state sequences $\sigma$ and all positions $s$. Let $\sigma$ be a state
  sequence such that $\sigma,s \models \ltl{Y^iX^i\psi}$ for a given $s \in
  \N$. We divide in cases:
  \begin{enumerate}[label=(\roman*)]
    \item
      if $s<i$, then $\sigma,s \not \models \ltl{Y^iX^i\psi}$, but this is
      a contradiction with our hypothesis;
    \item then it has to be the case that $s \ge i$. It holds that:
      \begin{align}
        &\sigma,s \models \ltl{Y^iX^i\psi} \\
        \Leftrightarrow \ &\sigma,s-i \models \ltl{X^i\psi} \\
        \Leftrightarrow \ &\sigma,s-i+i \models \psi \\
        \Leftrightarrow \ &\sigma,s \models \ltl{\psi \land Y^i \true} & \mbox{
          since } s \ge i
      \end{align}
  \end{enumerate}
  We prove the opposite direction, that is $\sigma,s \models \ltl{\psi \land
  Y^i\true}$ implies $\sigma,s \models \ltl{Y^iX^i\psi}$, for all state
  sequences $\sigma$ and all positions $s$. Let $\sigma$ be a state sequence
  such that $\sigma,s \models \ltl{\psi \land Y^i \true}$ for a given $s \in
  \N$. We divide in cases: 
  \begin{enumerate}[label=(\roman*)]
    \item
      if $s<i$, then $\sigma,s \not \models \ltl{Y^i\true}$, but this is
      a contradiction with our hypothesis;
    \item then it has to be the case that $s \ge i$. It holds that:
      \begin{align}
        &\sigma,s \models \ltl{\psi \land Y^i\true}  \\
        \Leftrightarrow \ &\sigma,s-i \models \ltl{X^i\psi} & \mbox{ since } s \ge i\\
        \Leftrightarrow \ &\sigma,s-i+i \models \ltl{Y^iX^i\psi} \\
        \Leftrightarrow \ &\sigma,s \models \ltl{Y^iX^i \psi}
      \end{align}
  \end{enumerate}
  This concludes the proof for $\bar{R_3}$.

  Consider now the $\bar{R_4}$ case. We first prove the left-to-right
  direction, that is $\sigma,s \models \ltl{Y^i(\psi_1 R \psi_2)}$ implies
  $\sigma,s \models \ltl{(Y^i\psi_1) R (Y^i\psi_2)}$, for all state sequences
  $\sigma$ and all positions $s$. Let $\sigma$ be a state sequence such that
  $\sigma,s \models \ltl{Y^i(\psi_1 R \psi_2)}$ with $s \ge i$ (obviously, it
  can't be that $s < i$). It holds that $\sigma, s-i \models \psi_1 R \psi_2$.
  Now, we divide in cases:
  \begin{enumerate}
    \item
      if $\forall k \ge s-i. \sigma,k \models \psi_2$, then $\forall k \ge s.
      \sigma,k \models \ltl{Y^i \psi_2}$ and thus $\sigma,s \models
      \ltl{(Y^i\psi_1) R (Y^i\psi_2)}$.
    \item 
      if $\exists k \ge s-i. (\sigma,k \models \psi_2 \land \forall s-i \le
      l \le k. \sigma,l \models \psi_1)$, then $\exists k \ge s. (\sigma,k
      \models \ltl{Y^i\psi_2} \land \forall s \le l \le k. \sigma,l \models
      \ltl{Y^i\psi_1})$, and thus $\sigma,s \models \ltl{(Y^i\psi_1)
      R (Y^i\psi_2)}$.
  \end{enumerate}
  Now we prove the opposite direction. Suppose that $\sigma,s \models \ltl{(Y^i
  \psi_1) R (Y^i \psi_2)}$ where $s \ge i$. We divide in cases:
  \begin{enumerate}
    \item
      if $\forall k \ge s. \sigma,k \models \ltl{Y^i\psi_2}$, then:
      \begin{align}
        &\forall k \ge s-i. \sigma,k \models \psi_2 \\
        \Leftrightarrow \ &\sigma,s-i \models \ltl{\psi_1 R \psi_2} \\
        \Leftrightarrow \ &\sigma,s \models \ltl{Y^i(\psi_1 R \psi_2)}
      \end{align}
    \item
      if $\exists k \ge s. (\sigma,k \models \ltl{Y^i\psi_1} \land \forall
      k \le l \le k. \sigma,l \models \ltl{Y^i\psi_2})$, then:
      \begin{align}
        &\exists k \ge s-i. (\sigma,k \models \psi_1 \land \forall s-i \le
        l \le k. \sigma,l \models \psi_2) \\
        \Leftrightarrow \ &\sigma,s-i \models \ltl{\psi_1 R \psi_2} \\
        \Leftrightarrow \ &\sigma,s \models \ltl{Y^i(\psi_1 R \psi_2)}
      \end{align}
  \end{enumerate}
  This concludes the proof for the $\bar{R_4}$ case. 
  
  The case for $\bar{R_5}$ is simple, and it consists in the following steps.
  For all state sequences $\sigma$ and for all positions $s$, it holds that:
  \begin{align}
    &\sigma,s \models \ltl{GG\psi} \\
    \Leftrightarrow \ &\forall i \ge s. \sigma,i \models \ltl{G\psi} \\
    \Leftrightarrow \ &\forall i \ge s. \forall j \ge i. \sigma,j \models \psi \\
    \Leftrightarrow \ &\forall i \ge s. \sigma ,i \models \psi \\
    \Leftrightarrow \ &\sigma,s \models \ltl{G \psi}
  \end{align}
  Consider the $\bar{R_6}$ strong equivalence. We first prove the left-to-right
  direction. Suppose that $\sigma,s \models \ltl{G(\psi_1 R \psi_2)}$, for
  a given state sequence $\sigma$ and a given position $s$. It holds that
  $\forall i \ge s. \sigma,i \models \ltl{\psi_1 R \psi_2}$. We divide in
  cases, depending on the semantics of the \emph{release} operator:
  \begin{enumerate}
    \item if $\forall i \ge s. \forall j \ge i. \sigma,j \models \psi_2$. In
      this case we have that $\forall i \ge s. \sigma,i \models \psi_2$, that
      is $\sigma,s \models \ltl{G\psi_2}$.
    \item otherwise, $\forall i \ge s. \exists j \ge i. ( \sigma,j \models \psi_1 \land
      \forall i \le k \le j. \sigma,k \models \psi_2)$. In particular, for
      $k=i$, we have that $\forall i \ge s. \sigma,i \models \psi_2$, that is
      $\sigma,s \models \ltl{G\psi_2}$.
  \end{enumerate}
  We prove the right-to-left direction for the $\bar{R_6}$ case. Suppose that
  $\sigma,s \models \ltl{G\psi_2}$, for a given state sequence $\sigma$ and
  position $s$. It holds that:
  \begin{align}
    &\sigma,s \models \ltl{G\psi_2} \\
    \Leftrightarrow \ &\forall i \ge s. \sigma,i \models \psi_2 \\
    \Leftrightarrow \ &\forall i \ge s. \forall j \ge i. \sigma,j \models
      \psi_2 \\
    \Rightarrow \ &\forall i \ge s. \sigma,i \models \ltl{\psi_1 R \psi_2} \\
    \Leftrightarrow \ &\sigma,s \models \ltl{G(\psi_1 R \psi_2)}
  \end{align}
  Finally, consider the case for the $\bar{R_7}$ strong equivalence. We first
  prove the left-to-right direction. Suppose that $\sigma,s \models \ltl{\psi_1
  R (G \psi_2)}$ for a given state sequence $\sigma$ and position $s$. We
  divide in cases, depending on the semantics of the \emph{release} operator:
  \begin{enumerate}
    \item if $\forall i \ge s. \sigma,i \models \ltl{G\psi_2}$, then for $i=s$
      we have that $\sigma,s \models \ltl{G\psi_2}$.
    \item otherwise, $\exists i \ge s.(\sigma,i \models \psi_1 \land \forall
      s \le j \le i . \sigma,j \models \ltl{G\psi_2})$. In particular, for
      $j=s$, $\sigma,s \models \ltl{G\psi_2}$.
  \end{enumerate}
  Therefore, in both cases we have that $\sigma,s \models \ltl{G\psi_2}$. For
  the right-to-left direction, suppose that $\sigma,s \models \ltl{G\psi_2}$.
  Then, $\forall i \ge s. \sigma,i \models \ltl{G\psi_2}$. This implies that
  $\sigma,s \models \ltl{\psi_1 R (G \psi_2)}$. This concludes the proof of
  all the auxiliary strong equivalences.

  We can now prove the remaining rules $R_3$-$R_7$. Consider first $R_3$ in the
  case $i>j$: we have to prove that $\sigma,s \models \ltl{(X^i \psi_1) R (X^j
  \psi_2) \ \iff \ \sigma,s \models X^i(\psi_1 R (Y^{i-j}\psi_2))}$, for all
  states sequences $\sigma$ and all positions $s$. This can be simply done by
  means of the auxiliary rules $\bar{R_2}$ and $\bar{R_3}$:
  \begin{align}
    &\sigma,s \models \ltl{(X^i \psi_1) R (X^j \psi_2)} \\
    \iff \ &\sigma,s \models \ltl{X^i(\psi_1 R (Y^i X^j \psi_2))} 
      & \mbox{by rule } \bar{R_2} \\
    \iff \ &\sigma,s \models \ltl{X^i(\psi_1 R (Y^{i-j}(Y^jX^j\psi_2)))} \\
    \iff \ &\sigma,s \models \ltl{X^i(\psi_1 R (Y^{i-j}(\psi_2 \land Y^j \true)))} 
      & \mbox{by rule } \bar{R_3} \\
    \iff \ &\sigma,s \models \ltl{X^i(\psi_1 R (Y^{i-j}\psi_2 \land Y^{i-j+j} \true))} \\
    \iff \ &\sigma,s \models \ltl{X^i(\psi_1 R (Y^{i-j}\psi_2 \land Y^{i} \true))} \\
    \iff \ &\sigma,s \models \ltl{X^i(\psi_1 R (Y^{i-j}\psi_2))}
  \end{align}
  Consider now the rule $R_3$ in the case $i \le j$. We have to prove that
  $\sigma,s \models \ltl{(X^i \psi_1) R (X^j \psi_2) \ \iff \ \sigma,s \models
  X^j((Y^{j-i}\psi_1) R \psi_2)}$.  This can be done using the auxiliary
  equivalences $\bar{R_1}$ and $\bar{R_3}$:
  \begin{align}
    &\sigma,s \models \ltl{(X^i \psi_1) R (X^j \psi_2)} \\
    \iff \ &\sigma,s \models \ltl{X^j((Y^jX^i\psi_1) R \psi_2)} 
      & \mbox{by rule } \bar{R_1} \\
    \iff \ &\sigma,s \models \ltl{X^j((Y^{j-i}(Y^iX^i\psi_1)) R \psi_2)} \\
    \iff \ &\sigma,s \models \ltl{X^j((Y^{j-i}(\psi_1 \land Y^i\true)) R \psi_2)} 
      & \mbox{by rule } \bar{R_3} \\
    \iff \ &\sigma,s \models \ltl{X^j((Y^{j-i}\psi_1 \land Y^{j-i+i}\true)R \psi_2)} \\
    \iff \ &\sigma,s \models \ltl{X^j((Y^{j-i}\psi_1 \land Y^j \true) R \psi_2)} \\
    \iff \ &\sigma,s \models \ltl{X^j((Y^{j-i}\psi_1) R \psi_2)}
  \end{align}
  Consider the $R_4$ rule in the case $i>j$. It holds that:
  \begin{align}
    &\sigma \models \ltl{(X^i \psi_1)R(X^j(\psi_2 R \psi_3))} \\
    \Leftrightarrow \ &\sigma,s \models \ltl{X^i(\psi_1R(Y^iX^j(\psi_2 R \psi_3)))}
      \\ & \mbox{by rule } \bar{R_2} \\
    \Leftrightarrow \ &\sigma,s \models \ltl{X^i(\psi_1R(Y^{i-j}Y^jX^j(\psi_2 R \psi_3)))} \\
    \Leftrightarrow \ &\sigma,s \models \ltl{X^i(\psi_1R(Y^{i-j}(\psi_2 R \psi_3 \land Y^j\true)))}
      \\ & \mbox{by rule } \bar{R_3} \\
    \Leftrightarrow \ &\sigma,s \models\ltl{X^i(\psi_1R(Y^{i-j}(\psi_2 R \psi_3) \land Y^i\true))} \\
    \Leftrightarrow \ &\sigma,s \models \ltl{X^i(\psi_1R(Y^{i-j}(\psi_2 R \psi_3))) \land 
      X^i(\psi_1 R Y^i\true)} 
      \\ & \mbox{by rule } R_1 \\
    \Leftrightarrow \ &\sigma,s \models \ltl{X^i(\psi_1R(Y^{i-j}(\psi_2 R \psi_3)))} \\ 
    \Leftrightarrow \ &\sigma,s \models \ltl{X^i(\psi_1R((Y^{i-j}\psi_2) R (Y^{i-j}\psi_3)))}
      \\ & \mbox{by rule } \bar{R_4}
  \end{align}
  Finally, consider the $R_4$ rule in the case $i \le j$. It holds that:
  \begin{align}
    &\sigma \models \ltl{(X^i \psi_1)R(X^j(\psi_2 R \psi_3))} \\
    \Leftrightarrow \ &\sigma,s \models \ltl{X^j((Y^jX^i \psi_1) R (\psi_2 R \psi_3))}
      & \mbox{by rule } \bar{R_1} \\
    \Leftrightarrow \ &\sigma,s \models \ltl{X^j((Y^{j-i}Y^iX^i \psi_1) R (\psi_2 R \psi_3))} \\
    \Leftrightarrow \ &\sigma,s \models \ltl{X^j((Y^{j-i}(\psi_1 \land Y^i \true)) R (\psi_2 R \psi_3))}
      & \mbox{by rule } \bar{R_3} \\
    \Leftrightarrow \ &\sigma,s \models \ltl{X^j((Y^{j-i}\psi_1 \land Y^j
      \true) R (\psi_2 R \psi_3))} \\
    \Leftrightarrow \ &\sigma,s \models \ltl{X^j((Y^{j-i}\psi_1) R (\psi_2 R \psi_3))}
  \end{align}
  Consider the $R_5$ rule. It can be proved by means of the rules $R_4$ and
  $\bar{R_5}$ as follows. For all state sequences $\sigma$ and all positions
  $s$, it holds that:
  \begin{align}
    &\sigma,s \models \ltl{GX^iG\psi} \\
    \Leftrightarrow \ &\sigma,s \models \ltl{(X^0 \false) R (X^i(\false R \psi))} 
      \\ & \mbox{ by definition of \emph{globally} operator} \\
    \Leftrightarrow \ &\sigma,s \models \ltl{ X^i( (Y^i \false) R ( \false R \psi ) ) }
      & \mbox{ by rule } R_4 \\
    \Leftrightarrow \ &\sigma,s \models \ltl{ X^i( \false R (\false R \psi) ) } \\
    \Leftrightarrow \ &\sigma,s \models \ltl{ X^i( G G \psi ) } \\
    \Leftrightarrow \ &\sigma,s \models \ltl{ X^i G \psi }
      & \mbox{ by rule } \bar{R_5}
  \end{align}
  Consider the $R_6$ rule. It can be prove by means of the rules $R_4$ and
  $\bar{R_6}$ as follows. For all state sequences $\sigma$ and positions $s$ it
  holds that:
  \begin{align}
    &\sigma,s \models \ltl{GX^i(\psi_1 R \psi_2)} \\
    \Leftrightarrow \ &\sigma,s \models \ltl{((X^0 \false) R (X^i (\psi_1 R \psi_2)))} 
      \\ & \mbox{ by definition of \emph{globally} operator} \\
    \Leftrightarrow \ &\sigma,s \models \ltl{ X^i( (Y^i \false) R (\psi_1 R \psi_2))}
      & \mbox{ by rule } R_4 \\
    \Leftrightarrow \ &\sigma,s \models \ltl{ X^i( \false R (\psi_1 R \psi_2))} \\
    \Leftrightarrow \ &\sigma,s \models \ltl{ X^i( G (\psi_1 R \psi_2))} \\
    \Leftrightarrow \ &\sigma,s \models \ltl{ X^i( G \psi_2 )}
      & \mbox{ by rule } \bar{R_6}
  \end{align}
  Consider the $R_7$ rule. It can be proved by means of the rules $R_4$ and
  $\bar{R_7}$ as follows. Let $\sigma$ be a state sequence and let $s$ be
  a position. We divide in cases. If $i>j$, then:
  \begin{align}
    &\sigma,s \models \ltl{(X^i\psi_1) R (X^jG\psi_2)} \\
    \Leftrightarrow \ &\sigma,s \models \ltl{(X^i\psi_1) R (X^j(\false R \psi_2))} \\
    \Leftrightarrow \ &\sigma,s \models \ltl{X^i(\psi_1 R ((Y^{i-j}\false) R (Y^{i-j}\psi_2)))}
      & \mbox{ by rule } R_4 \\
    \Leftrightarrow \ &\sigma,s \models \ltl{X^i(\psi_1 R (\false R (Y^{i-j}\psi_2)))} \\
    \Leftrightarrow \ &\sigma,s \models \ltl{X^i(\psi_1 R (G (Y^{i-j}\psi_2)))} \\
    \Leftrightarrow \ &\sigma,s \models \ltl{X^i(\psi_1 R (G (Y^{i-j}\psi_2)))} 
      &\mbox{ by rule } R_7 \\
    \Leftrightarrow \ &\sigma,s \models \ltl{X^i G Y^{i-j}\psi_2} 
  \end{align}
  Otherwise, it holds that $i \le j$ and:
  \begin{align}
    &\sigma,s \models \ltl{(X^i\psi_1) R (X^jG\psi_2)} \\
    \Leftrightarrow \ &\sigma,s \models \ltl{(X^i\psi_1) R (X^j(\false R \psi_2))} \\
    \Leftrightarrow \ &\sigma,s \models \ltl{X^j((Y^{j-i}\psi_1) R (\false R \psi_2))}
      & \mbox{ by rule } R_4 \\
    \Leftrightarrow \ &\sigma,s \models \ltl{X^j((Y^{j-i}\psi_1) R (G \psi_2))} \\
    \Leftrightarrow \ &\sigma,s \models \ltl{X^j G \psi_2} 
      & \mbox{ by rule } \bar{R_7}
  \end{align}
  This concludes the case for the rules $R_1$-$R_7$.
  
  It remains the case for the $R_{flat}$ rule, for which we have to prove only
  equivalence. We first prove the left-to-right direction, for all $n \ge 3$.
  Suppose that:
  \begin{align}
    &\sigma,0 \models \ltl{X^i( \psi_1 R (\psi_2 R ( \dots (\psi_{n-1} R \psi_n)
      \dots )) )} \\
    &\sigma,i \models \ltl{ \psi_1 R (\psi_2 R ( \dots (\psi_{n-1} R \psi_n)
      \dots ))}
  \end{align}
  This formula contains exactly $n$ \emph{release} operators. Each of these can
  be satisfied in two ways:
  \begin{enumerate*}[label=(\roman*)]
    \item \emph{universally}, that is if for all the future positions the
      right-hand side formula holds, or
    \item \emph{existentially}, if there exists a position in the future where
      the left-hand side formula holds and the right-hand side formula holds
      until then.
  \end{enumerate*}
  Therefore, we have a total of $2^{n-1}$ cases.

  We consider first the cases in which there exists a \emph{release} operator
  that is universally satisfied. These correspond to $2^{n-1}-1$ cases. Let $m$
  be the index of the outermost between these operators. Let $k_1 = i$. We have
  that:
  \begin{align}
    &\exists j_1 \ge k_1. (
      \sigma,j_1 \models \psi_1 \land \forall k_1 \le k_2 \le j_1. \\
        &\exists j_2 \ge k_2. ( \sigma,j_2 \models \psi_2 \land
          \dots \land \forall k_{m-1} \le k_{m-1} \le j_{m-2} . \\
        &\forall k_m \ge k_{m-1}.
          (\sigma,k_m \models \ltl{\psi_m R (\dots (\psi_{n-1} R \psi_n)\dots)})
        )\dots
    )
  \end{align}
  Which is equivalent to:
  \begin{align}
    &\exists j_1 \ge k_1. (
      \sigma,j_1 \models \psi_1 \land \forall k_1 \le k_2 \le j_1. \\
        &\exists j_2 \ge k_2. ( \sigma,j_2 \models \psi_2 \land
          \dots \land \forall k_{m-1} \le k_{m-1} \le j_{m-2} . \\
        &(\sigma,k_{m-1} \models \ltl{G (\psi_m R (\dots (\psi_{n-1} R \psi_n) \dots))}
        )\dots
    ))
  \end{align}
  By the repeated application of the $\bar{R_6}$ auxiliary rule $n-m$ times, we
  have that:
  \begin{align}
    &\exists j_1 \ge k_1. (
      \sigma,j_1 \models \psi_1 \land \forall k_1 \le k_2 \le j_1. \\
        &\exists j_2 \ge k_2. ( \sigma,j_2 \models \psi_2 \land
          \dots \land \forall k_{m-1} \le k_{m-1} \le j_{m-2} . \\
        &(\sigma,k_{m-1} \models \ltl{G \psi_n}
        )\dots
    ))
  \end{align}
  that is:
  \begin{align}
    &\exists j_1 \ge k_1. (
      \sigma,j_1 \models \psi_1 \land \forall k_1 \le k_2 \le j_1. \\
        &\exists j_2 \ge k_2. ( \sigma,j_2 \models \psi_2 \land
          \dots \land \forall k_{m-1} \le k_{m-1} \le j_{m-2} . \\
        &\forall k \ge k_{m-1}.(\sigma,k \models \psi_n
        )\dots
    ))
  \end{align}
  In particular, for $k_1 = k_2 = \dots = k_{m-2} = k_{m-1}$, we have that:
  \begin{align}
    \forall k \ge k_1. \sigma,k \models \psi_n
  \end{align}
  Since by definition $k_1 = i$, we have that $\forall k \ge i. \sigma,k
  \models \psi_n$, and thus $\sigma,0 \models \ltl{X^i( (\psi_{n-1} \land
  O (\psi_{n-2} \land \dots O(\psi_1 \land Y^i \true))) R \psi_n )}$.
  The remaining case is when \emph{all} the \emph{release} operators are
  existentially satisfied. Suppose that:
  \begin{align}
    &\exists j_1 \ge k_1. (
      \sigma,j_1 \models \psi_1 \land \forall k_1 \le k_2 \le j_1. \\
    &\exists j_2 \ge k_2. ( \sigma,j_2 \models \psi_2 \land
      \dots \land \forall k_{n-1} \le k_{n-1} \le j_{n-2} . \\
    &\exists j_{n-1} \ge k_{n-1}. (
      \sigma, j_{n-1} \models \psi_{n-1} \land
      \forall k_{n-1} \le k_n \le j_{n-1}. \\
      &\quad\sigma,k_n \models \psi_n
    )
    )\dots
    )
  \end{align}
  where $k_1 = i$. This implies that:
  \begin{align}
    &\exists j_1 \ge i. ( \sigma,j_1 \models \psi_1 \land \\
    &\exists j_2 \ge j_1. ( \sigma,j_2 \models \psi_2 \land \dots \land \\
    &\exists j_{n-1} \ge j_{n-2}. (
      \sigma, j_{n-1} \models \psi_{n-1} \land 
      \forall i \le k \le j_{n-1}. \\
      &\quad\sigma,k \models \psi_n
    )
    \dots
    ))
  \end{align}
  This is equivalent to:
  \begin{align}
    &\exists j_{n-1} \ge i. ( \sigma,j_{n-1} \models \psi_{n-1} \land \\
    &\exists i \le j_{n-2} \le j_{n-1}. ( \sigma,j_{n-2} \models \psi_{n-2} \land \dots \land \\
    &\exists i \le j_1 \le j_2. ( \sigma, j_1 \models \psi_1 ) \dots) \land \\ 
    &\quad\forall i \le k \le j_{n-1}. \sigma,k \models \psi_n)
  \end{align}
  This in turn is equivalent to:
  \begin{align}
    &\exists j_{n-1} \ge i. ( \sigma,j_{n-1} \models \psi_{n-1} \land \\
    &\exists 0 \le j_{n-2} \le j_{n-1}. ( \sigma,j_{n-2} \models \psi_{n-2} \land \dots \land \\
    &\exists 0 \le j_1 \le j_2. ( \sigma, j_1 \models \ltl{\psi_1 \land Y^i \true}) \dots) \land \\ 
    &\quad\forall i \le k \le j_{n-1}. \sigma,k \models \psi_n)
  \end{align}
  This is the definition of the existential semantics of the formula
  $\ltl{(\psi_{n-1} \land O(\psi_{n-2} \land \dots O(\psi_1 \land Y^i \true)))
  R \psi_n}$, starting from position $i$. Therefore, $\sigma,0 \models \ltl{
    X^i( (\psi_{n-1} \land O(\psi_{n-2} \land \dots O(\psi_1 \land Y^i \true)))
    R \psi_n) }$.

  We now prove the right-to-left direction for $R_{flat}$. Suppose that
  $\sigma,0 \models \ltl{X^i( (\psi_{n-1} \land O(\psi_{n-2} \land \dots
  O(\psi_1 \land Y^i \true))) R \psi_n )}$.  Therefore, $\sigma,i \models
  \ltl{(\psi_{n-1} \land O(\psi_{n-2} \land \dots O(\psi_1 \land Y^i \true)))
  R \psi_n}$.
  We divide in cases:
  \begin{enumerate}
    \item
      if $\forall j \ge i. \ \sigma,j \models \psi_n$, then \\$\sigma,0 \models
      \ltl{X^i( \psi_1 R(\psi_2 R(\dots (\psi_{n-1} R \psi_n) \dots)) )}$
    \item otherwise, $\exists j \ge i.( \sigma,j \models \ltl{\psi_{n-1} \land
      O(\psi_{n-2} \land \dots O(\psi_1 \land Y^i \true)\dots)} \land \forall
      i \le k \le j. \sigma,k \models \psi_n)$.
  \end{enumerate}
  With the former case, we are done. Instead, the latter is equivalent to:
  \begin{align}
    &\exists j_{n-1} \ge i. ( \sigma,j_{n-1} \models \psi_{n-1} \land \\
    &\exists 0 \le j_{n-2} \le j_{n-1}.( \sigma,j_{n-2} \models \psi_{n-2} \land \dots \\
    &\exists 0 \le j_1 \le j_2.(\sigma,j_1 \models (\psi_1 \land \ltl{Y^i \true}) )
    \dots ) \land \\
    &\forall i \le k \le j_{n-1}. \sigma,k \models \psi_n
    )
  \end{align}
  In turn, this is equivalent to:
  \begin{align}
    &\exists j_{n-1} \ge i. ( \sigma,j_{n-1} \models \psi_{n-1} \land \\
    &\exists i \le j_{n-2} \le j_{n-1}.( \sigma,j_{n-2} \models \psi_{n-2} \land \dots \\
    &\exists i \le j_1 \le j_2.(\sigma,j_1 \models \psi_1  )
    \dots) \land \\
    &\forall i \le k \le j_{n-1}. \sigma,k \models \psi_n
    )
  \end{align}
  This is equivalent to:
  \begin{align}
    &\exists j_{1} \ge i. ( \sigma,j_{1} \models \psi_{1} \land \\
    &\exists j_2 \ge j_1. ( \sigma,j_{2} \models \psi_{2} \land \dots \\
    &\exists j_{n-1} \ge j_{n-2}. (\sigma,j_{n-1} \models \psi_{n-1}  )
    \dots ) \land \\
    &\forall i \le k \le j_{1}. \sigma,k \models \psi_n
    )
  \end{align}
  which implies that:
  \begin{align}
    &\exists j_{1} \ge i. ( \sigma,j_{1} \models \psi_{1} \land \forall i \le k_1 \le j_1. \\
    &\exists j_2 \ge j_1. ( \sigma,j_{2} \models \psi_{2} \land \dots \land
      \forall k_{n-2} \le k_{n-1} \le j_{n-1} . \\
    &\exists j_{n-1} \ge j_{n-2}. (\sigma,j_{n-1} \models \psi_{n-1} \land 
      \forall k_{n-1} \le k \le j_{n-1}. \\
    &\qquad\sigma,k \models \psi_n
    ) \dots
    ))
  \end{align}
  This is the definition of the \emph{existential} semantics of the formula
  $\ltl{\psi_1 R (\psi_2 R (\dots (\psi_{n-1} R \psi_n) \dots))}$, starting
  from position $i$. Therefore, $\sigma,0 \models \ltl{X^i( \psi_1 R (\psi_2
  R (\dots (\psi_{n-1} R \psi_n) \dots)) )}$.
  This concludes the proof of \cref{app:lemma:equivalences}.
\end{IEEEproof}

\begin{lemma}
\label{app:lemma:subroutines}
  Let $\psi_1$, $\psi_2$  and $\psi_3$ be \LTLFP formulas. Let $\phi$ be
  a formula of type $\ltl{X^j\psi_2}$, $\ltl{X^jG\psi_2}$ or $\ltl{X^j(\psi_2
  R \psi_3)}$. For each state sequence $\sigma$ and position $i$, it holds
  that: 
  \begin{enumerate}
    \item 
      $\sigma,i \models \ltl{G\phi} \ \iff \ \sigma,i \models
      \texttt{resolve\_globally}(\phi)$
    \item
      $\sigma,i \models \ltl{(X^i\psi_1)R\phi} \ \iff \\ \sigma,i \models
      \texttt{resolve\_release}(\ltl{X^i\psi_1, \phi})$
  \end{enumerate}
\end{lemma}
\begin{IEEEproof}
  We prove the second point, for the \emph{release} operator.
  The subroutine \texttt{resolve\_release} divides in cases, depending on the
  structure of $\phi$:
  \begin{itemize}
    \item
      if $\phi = \ltl{X^j\psi_2}$ and $i>j$, then:
      \begin{align}
        \texttt{resolve\_release}(\ltl{X^i\psi_1,X^j\psi_2}) \coloneqq \\
        \ltl{X^i(\psi_1 R (Y^{i-j}\psi_2 ))} 
      \end{align}
      By rule $R_3$ of \cref{app:lemma:equivalences}, we have that $\sigma,i \models
      \ltl{(X^i\psi_1)R\phi} \iff \sigma,i \models
      \texttt{resolve\_release}(\ltl{X^i\psi_1, \phi})$.
    \item
      if $\phi = \ltl{X^j\psi_2}$ and $i\le j$, then
      \begin{align}
        \texttt{resolve\_release}(\ltl{X^i\psi_1,X^j\psi_2}) \coloneqq \\
        \ltl{X^j((Y^{j-i}\psi_1)R\psi_2)}
      \end{align}
      By rule $R_3$ of \cref{app:lemma:equivalences}, we have that $\sigma,i \models
      \ltl{(X^i\psi_1)R\phi} \iff \sigma,i \models
      \texttt{resolve\_release}(\ltl{X^i\psi_1, \phi})$.
    \item
      if $\phi = \ltl{X^j(\psi_2 R \psi_3)}$ and $i>j$, then
      \begin{align}
        \texttt{resolve\_release}(\ltl{X^i\psi_1,X^j(\psi_2 R \psi_3)})
        \coloneqq \\
        \ltl{X^i( \psi_1 R ((Y^{i-j}\psi_2) R (Y^{i-j} \psi_3) ) )}
      \end{align}
      By rule $R_4$ of \cref{app:lemma:equivalences}, we have that $\sigma,i \models
      \ltl{(X^i\psi_1)R\phi} \iff \sigma,i \models
      \texttt{resolve\_release}(\ltl{X^i\psi_1, \phi})$.
    \item
      if $\phi = \ltl{X^j(\psi_2 R \psi_3)}$ and $i\le j$, then
      \begin{align}
        \texttt{resolve\_release}(\ltl{X^i\psi_1,X^j(\psi_2 R \psi_3)})
        \coloneqq \\
        \ltl{X^j( (Y^{j-i}\psi_1 ) R (\psi_2 R \psi_3) )}
      \end{align}
      By rule $R_4$ of \cref{app:lemma:equivalences}, we have that $\sigma,i \models
      \ltl{(X^i\psi_1)R\phi} \iff \sigma,i \models
      \texttt{resolve\_release}(\ltl{X^i\psi_1, \phi})$.
    \item
      if $\phi = \ltl{X^j G \psi_2 }$ and $i>j$, then
      \begin{align}
        \texttt{resolve\_release}(\ltl{X^i\psi_1,X^j G \psi_2 })
        \coloneqq \\
        \ltl{X^i G Y^{i-j} \psi_2}
      \end{align}
      By rule $R_7$ of \cref{app:lemma:equivalences}, we have that $\sigma,i \models
      \ltl{(X^i\psi_1)R\phi} \iff \sigma,i \models
      \texttt{resolve\_release}(\ltl{X^i\psi_1, \phi})$.
    \item
      if $\phi = \ltl{X^j G \psi_2 }$ and $i\le j$, then
      \begin{align}
        \texttt{resolve\_release}(\ltl{X^i\psi_1,X^j G \psi_2 })
        \coloneqq \\
        \ltl{X^j G \psi_2}
      \end{align}
      By rule $R_7$ of \cref{app:lemma:equivalences}, we have that $\sigma,i \models
      \ltl{(X^i\psi_1)R\phi} \iff \sigma,i \models
      \texttt{resolve\_release}(\ltl{X^i\psi_1, \phi})$.
  \end{itemize}
  The case for $\texttt{resolve\_globally}(\phi)$ is analogous. 
\end{IEEEproof}

\begin{lemma}[Soundness of $\apply(\cdot)$]
\label{app:lemma:applysound}
  For any \PLTLEBR formula $\phi$, for any state sequence $\sigma$ and for any
  position $i$, it holds that $\sigma,i \models \phi$ iff $\sigma,i \models
  \apply(\phi)$.
\end{lemma}
\begin{IEEEproof}
  Consider the pseudo-code of $\apply(\cdot)$ as described in \cref{app:fig:applyalg}. We
  prove this claim by induction on the complexity of formula $\phi$.
  
  The base case corresponds to the case when $\phi$ is a \LTLFP formula. In
  this case, the $\apply(\cdot)$ algorithm returns $\phi$ it self. Obviously,
  $\phi$ is strongly equivalent to $\apply(\phi)$

  For the inductive step, we divide in cases. If $\ltl{\phi := X\phi_1}$, then
  $\sigma,i+1 \models \phi_1$.  By inductive hypothesis $\sigma^\prime,i^\prime
  \models \phi_1$ iff $\sigma^\prime,i^\prime \models \apply(\phi_1)$, for all
  state sequences $\sigma^\prime$ and positions $i^\prime$. Therefore:
  \begin{align}
    \sigma,i \models \ltl{X\phi_1} \Leftrightarrow &\sigma,i+1 \models \phi_1 \\ 
    \Leftrightarrow &\sigma,i+1 \models \apply(\phi_1) \\
      &\quad\mbox{by inductive hypothesis} \\
    \Leftrightarrow &\sigma,i \models \ltl{X (\apply(\phi_1))}
  \end{align}
  In general, $\apply(\phi_1)$ is a conjunction of formulas of
  type $\ltl{X^j\psi}$, $\ltl{X^jG\psi}$, $\ltl{X^j((X^k\psi_1) R \psi_2)}$,
  that is:
  \begin{align}
    \apply(\phi_1) \coloneqq \phi^c_2 \land \dots \land \phi^c_n
  \end{align}
  and thus:
  \begin{align}
    \sigma,i \models \ltl{X\phi_1} \Leftrightarrow &\sigma,i \models \ltl{X
    (\phi^c_2 \land \dots \land \phi^c_n)}
  \end{align}
  Using rule $R_1$ of \cref{app:lemma:equivalences}, we have that:
  \begin{align}
    \sigma,i \models \ltl{X\phi_1} &\Leftrightarrow \sigma,i \models \ltl{X
    (\phi^c_2 \land \dots \land \phi^c_n)} \\
    &\Leftrightarrow \sigma,i \models
      \ltl{X\phi^c_2 \land \dots \land X\phi^c_n} \\
      &\qquad\mbox{by rule } R_1 \mbox{ of \cref{app:lemma:equivalences}} \\
    \sigma,i \models \phi &\Leftrightarrow \sigma,i \models \apply(\phi)
  \end{align}
  This concludes the case for $\phi \coloneqq \ltl{X\phi_1}$. Consider the case
  $\ltl{\phi := (X^i\psi_1)R\phi_1}$. Since by inductive hypothesis
  $\sigma^\prime, i^\prime \models \phi_1$ iff $\sigma^\prime, i^\prime \models
  \apply(\phi_1)$, for all state sequences $\sigma^\prime$ and positions
  $i^\prime$, we have that:
  \begin{align}
    \sigma,i \models \ltl{(X^i\psi_1)R\phi_1} \Leftrightarrow 
    &\sigma,i \models \ltl{(X^i\psi_1)R(\apply(\phi_1))} \\
    &\sigma,i \models \ltl{(X^i\psi_1)R(\phi^c_2 \land \dots \land \phi^c_n)}
  \end{align}
  where $\phi^c_i$ is a formula of type $\ltl{X^j\psi}$, $\ltl{X^jG\psi}$,
  $\ltl{X^j((X^k\psi_1) R \psi_2)}$, for each $1 < i \le n$.  By rule $R_2$ of
  \cref{app:lemma:equivalences}, we have that:
  \begin{align}
    \sigma,i \models \ltl{(X^i\psi_1)R\phi_1} &\Leftrightarrow \sigma,i \models
      \ltl{(X^i\psi_1)R(\phi^c_2 \land \dots \land \phi^c_n)} \\ 
    &\Leftrightarrow \sigma,i \models \ltl{(X^i\psi_1)R(\phi^c_2)} 
      \land \dots \land\\
    &\quad\qquad\qquad\ltl{(X^i\psi_1)R(\phi^c_n)}
  \end{align}
  Let $\phi^r_i \equiv \texttt{resolve\_release}(\ltl{X^i\psi_1},\phi^c_i)$,
  for all $1 < i \le n$.  By \cref{app:lemma:subroutines}:
  \begin{align}
    \sigma,i \models \ltl{(X^i\psi_1)R\phi_1} &\Leftrightarrow \sigma,i \models
      \ltl{(X^i\psi_1)R(\phi^c_2)} \land \dots \land\\
      &\quad\qquad\qquad\ltl{(X^i\psi_1)R(\phi^c_n)} \\
    &\Leftrightarrow \sigma,i \models 
      \ltl{(X^i\psi_1)R(\phi^r_2)} \land \dots \land\\
      &\quad\qquad\qquad\ltl{(X^i\psi_1)R(\phi^r_n)} \\
      &\qquad \mbox{by \cref{app:lemma:subroutines}} \\
    &\Leftrightarrow \sigma,i \models \apply(\phi) \\
      &\qquad\mbox{by definition of \apply}
  \end{align}
  This concludes the case for $\phi \coloneqq \ltl{\phi :=
  (X^i\psi_1)R\phi_1}$.  The case for the \emph{globally} operator is analogous
  to the proof for the \emph{release} one.
\end{IEEEproof}

\begin{lemma}[Soundness of $\flatten(\cdot)$]
\label{app:lemma:flattensound}
  For any \PLTLEBR formula $\phi$, it holds that $\phi \equiv \flatten(\phi)$.
\end{lemma}
\begin{IEEEproof}
  We prove this lemma by induction on the number $n$ of top-level conjucts or
  disjuncts. The base case corresponds to the case of $n=0$. We divide in
  cases:
  \begin{itemize}
    \item
      if $\phi \coloneqq \ltl{X^i(\psi_1 R (\psi_2 R (\dots (\psi_{n-1}
      R \psi_n)\dots)))}$, then $\flatten(\phi) \coloneqq \ltl{X^i( (\psi_{n-1}
      \land O(\psi_{n-2} \land \dots O(\psi_1 \land Y^i \true)\dots)) R \psi_n
      )}$. By the $R_{flat}$ rule of \cref{app:lemma:equivalences}, $\phi \equiv \flatten(\phi)$.
    \item 
      otherwise, the \flatten algorithm falls in the \texttt{default} case.
      In this case, $\flatten(\phi) \coloneqq \phi$, and obviously $\phi \equiv
      \flatten(\phi)$.
  \end{itemize}
  For the inductive step, we divide in cases as well.
  \begin{itemize}
    \item
      if $\phi \coloneqq \phi_1 \land \phi_2$, then by inductive hypothesis
      $\phi_1 \equiv \flatten(\phi_1)$ and $\phi_2 \equiv \flatten(\phi_2)$.
      Thus $\phi \equiv \flatten(\phi_1) \land \flatten(\phi_2)$, that is $\phi
      \equiv \flatten(\phi)$.
    \item 
      if $\phi \coloneqq \phi_1 \land \phi_2$, then by inductive hypothesis
      $\phi_1 \equiv \flatten(\phi_1)$ and $\phi_2 \equiv \flatten(\phi_2)$.
      Thus $\phi \equiv \flatten(\phi_1) \lor \flatten(\phi_2)$, that is $\phi
      \equiv \flatten(\phi)$.
  \end{itemize}
\end{IEEEproof}

\begin{lemma}[Soundness of $\canonize(\cdot)$]
\label{app:lemma:canonizesound}
  For any \PLTLEBR formula $\phi$, it holds that $\phi$ and $\canonize(\phi)$
  are equivalent and $\canonize(\phi)$ is a Canonical \PLTLEBR formula.
\end{lemma}
\begin{IEEEproof}
\label{app:proof:canonizeeq}
  We define $\canonize(\phi)$ as the formula $\flatten(\apply(\phi))$, where
  \apply is the algorithm in \cref{app:fig:applyalg} and \flatten is the
  algorithm in \cref{app:fig:flatten}.  By \cref{app:lemma:applysound}, for
  each state sequence $\sigma$ and position $i$, we have that $\sigma,i \models
  \phi$ iff $\sigma,i \models \apply(\phi)$. In particular, for $i=0$, this
  means that $\phi \equiv \apply(\phi)$.  By \cref{app:lemma:flattensound}, we
  have that $\flatten(\apply(\phi)) \equiv \apply(\phi)$, and thus $\phi \equiv
  \flatten(\apply(\phi))$, and by definition $\phi \equiv \canonize(\phi)$.
  
  Finally, it is easy to see that all the rules of
  \cref{app:lemma:equivalences}, except for $R_4$, replace a formula with a one
  in Canonical \PLTLEBR. Thus $\canonize(\phi)$ would be a Canonical \PLTLEBR
  formula if we did not consider the nested \emph{release} operators. Since
  this is exactly the case solved by the $R_{flat}$ rule and thus by the
  \flatten algorithm (which produces a formula in canonical form), we have that
  $\flatten(\apply(\phi))$, which by definition is $\canonize(\phi)$, is in
  Canonical \PLTLEBR.
\end{IEEEproof}

\begin{proposition}[Complexity of $\canonize(\cdot)$]
\label{app:prop:canonizesize}
  For any \PLTLEBR formula $\phi$, $\canonize(\phi)$ can be built in
  $\mathcal{O}(n)$ time, and the size of $\canonize(\phi)$ is $\mathcal{O}(n)$, 
  where $n = |\phi|$.
\end{proposition}
\begin{IEEEproof}
\label{app:proof:canonizesize}
  Since $\canonize(\phi) \coloneqq \flatten(\apply(\phi))$, we study the
  complexity of both \apply and \flatten.  At each iteration, algorithm
  $\apply(\phi)$ makes at most one recursive call on a formula $\phi^\prime$ of
  size $|\phi^\prime| < |\phi|$ and thus it stop at most after $\mathcal{O}(n)$
  iterations. The same holds for \flatten. At each iteration, \apply and
  \flatten produce a formula of constant size with respect to the size of the
  formula produced by the recursive call; therefore the recurrence equation
  describing the size of the formula produced by $\canonize(\phi)$ is:
  \begin{equation}
    S(n)=
    \begin{cases}
      \mathcal{O}(1) & \text{if}\ n=1 \\
      S(n-1) + \mathcal{O}(1) & \text{otherwise}
    \end{cases}
  \end{equation}
  Therefore:
  \begin{align}
    S(n) &= S(n-1-i) + i \cdot \mathcal{O}(i) \\
    &= S(1) + \mathcal{O}(n) & \text{for } i=n-2 \\
    &\in \mathcal{O}(n)
  \end{align}
\end{IEEEproof}

\begin{lemma}
\label{app:lemma:pastltlebr}
  For each canonical \PLTLEBR formula $\phi$, for each \LTLFP formula $\alpha
  \in \LTLFP$ and for each $i \ge 0$, $\sigma(i) \models \alpha$ iff $\tau(i)
  \models v_\alpha$, where $\tau$ is the trace of $\autom(\phi)$ induced by
  $\sigma$. 
\end{lemma}
\begin{IEEEproof}
\label{app:proof:pastltlebr}
  We prove the lemma by induction on the structure of $\alpha$. For the base
  case, $\sigma(i) \models p \in \Sigma$ iff $\tau(i) \models v_p$; since by
  definition of its monitor $v_p \iff p$, we have that $\sigma(i) \models p$
  iff $\tau(i) \models p$; since $\tau$ is induced by $\sigma$, this is always
  true.

  For the inductive step, consider first $\alpha \lor \beta$. If $\sigma(i)
  \models \alpha \lor \beta$, then either $\sigma(i) \models \alpha$ or
  $\sigma(i) \models \beta$; by inductive hypothesis, either $\tau(i) \models
  v_\alpha$ or $\tau(i) \models v_\beta$; finally, by the definition of the
  monitor for disjunction, we have that $\tau(i) \models v_{\alpha \lor
  \beta}$.  The opposite case and the case for $\lnot \alpha$ can be proved
  similarly.

  Consider the case for $\ltl{Y\alpha}$. If $\sigma(i) \models \ltl{Y\alpha}$,
  then $\sigma(i-1) \models \alpha$ and $i>0$. By inductive hypothesis
  $\tau(i-1) \models v_\alpha$ and $i>0$; by definition of the monitor for
  $\ltl{Y\alpha}$, $\tau(i) \models v_{\ltl{Y\alpha}}$.

  Finally, we prove the case for $\ltl{\alpha S \beta}$. If $\sigma(i) \models
  \ltl{\alpha S \beta}$, then either $\sigma(i) \models \beta$ or $\sigma(i)
  \models \alpha \land \ltl{Y(\alpha S \beta)}$; by inductive hypothesis,
  either $\tau(i) \models v_\beta$ or $\tau(i) \models v_\alpha \land
  v_{\ltl{Y(\alpha S \beta)}}$; by definition of the monitor for $\ltl{\alpha
  S \beta}$, we have that $\tau(i) \models v_{\alpha S \beta}$. The opposite
  direction can be proved in the specular way.
\end{IEEEproof}

\begin{proposition}
\label{prop:dssasize}
  Let  $\phi$ be a canonical \PLTLEBR formula, with $|\phi| = n$. Then, there
  exists a deterministic SSA of size $\mathcal{O}(n)$ that accepts the same
  language.
\end{proposition}
\begin{IEEEproof}
  Let $\phi$ be a canonical \PLTLEBR formula over the alphabet $\Sigma$ and let
  $\autom(\phi) = (X \cup \Sigma, I(X), T(X,\Sigma,X^\prime), G(X))$ be the
  deterministic symbolic safety automaton as previously defined.
  
  \emph{Soundness.} We first prove that $\lang(\phi) = \lang(\autom(\phi))$. In
  particular we prove that $\forall \sigma \in \lang(\phi). \sigma \models
  \phi$ iff $\tau(i) \models S(X) \ \forall i \ge 0$, where $\tau$ is the trace
  induced by $\sigma$ in $\autom(\phi)$. Recall that $S(X) = \phi[\varphi
  / \lnot error_{\varphi}]$.  We proceed by induction on the structure of
  $\phi$.

  For the base case we consider $\phi = \ltl{X^iG\alpha}$ where $\alpha \in
  \LTLFP$(the cases for $\ltl{X^i\alpha}$ and $\ltl{X^i(\alpha R \beta)}$ are
  similar). If $\sigma \models \ltl{X^iG\alpha}$ then $\sigma(i) \models
  \ltl{G\alpha}$, that is $\sigma(j) \models \alpha \ \forall j \ge i$. By
  \cref{app:lemma:pastltlebr}, $\tau(j) \models v_\alpha \ \forall j \ge i$. The
  following points hold:
    \begin{enumerate}
      \item given the first condition in the monitor for $\ltl{X^iG\alpha}$, we
        have that $\tau(j) \models \lnot error_\phi \ \forall 0 \le j < i$;
      \item given the previous point and the fact that $\tau(j) \models
        v_\alpha \ \forall j \ge i$, by the second condition of the monitor we
        have that $\tau(j) \models \lnot error_\phi \ \forall j \ge i$.
    \end{enumerate}
  By these two points, it follows that $\tau(j) \models \lnot error_\phi \ 
  \forall j \ge 0$. Viceversa, if $\tau(j) \models \lnot error_\phi \ \forall
  j \ge 0$, then by definition of the monitor we have that $\tau(j) \models
  v_\alpha \ \forall j \ge i$. By \cref{app:lemma:pastltlebr}, $\sigma(j) \models
  \alpha \ \forall j \ge i$, that is $\sigma \models \ltl{X^iG\alpha}$.

  For the inductive step, consider first $\phi = \phi_1 \land \phi_2$. If
  $\sigma \models \phi$, then $\sigma \models \phi_1$ and $\sigma \models
  \phi_2$. By inductive hypothesis, $\tau(i) \models \phi_1[\varphi/\lnot
  error_\varphi] \ \forall i \ge 0$ and $\tau(i) \models \phi_2[\varphi / \lnot
  error_\varphi] \ \forall i \ge 0$, that is $\tau(i) \models (\phi_1 \land
  \phi_2)[\varphi / \lnot error_\varphi] \ \forall i \ge 0$. The opposite
  direction can be proved in the same way.

  Finally, consider the case $\phi = \phi_1 \lor \phi_2$. If $\sigma \models
  \phi$, then by inductive hypothesis either $\tau(i) \models \phi_1[\varphi
  / \lnot error_\varphi] \ \forall i \ge 0$ or $\tau(i) \models \phi_2[\varphi
  / \lnot error_\varphi] \ \forall i \ge 0$; thus $\tau(i) \models (\phi_1 \lor
  \phi_2)[\varphi / \lnot error_\varphi] \ \forall i \ge 0$. For the opposite
  direction, assume that $\tau(i) \models (\phi_1 \lor \phi_2)[\varphi / \lnot
  error_\varphi] \ \forall i \ge 0$; since each $error_\varphi$ is monotone (once
  set to true, it remains true forever), it holds that either $\tau(i) \models
  \phi_1[\varphi / \lnot error_\varphi] \ \forall i \ge 0$ or $\tau(i) \models
  \phi_2[\varphi / \lnot error_\varphi] \ \forall i \ge 0$. By inductive
  hypothesis, either $\sigma \models \phi_1$ or $\sigma \models \phi_2$, that
  is $\sigma \models \phi_1 \lor \phi_2$.

  \emph{Complexity.} Let $n = |\phi|$; it holds that:
  \begin{itemize}
    \item $|X| = |M_P|+|M_F| \in \mathcal{O}(n)$, since $|M_P|+|M_F| \le n$;
    \item $|I(X)|,|T(X,\Sigma,X^\prime)| \in \mathcal{O}(n)$, since they are
      both summations over the variables in $X$; 
    \item $|S(X)| \in \mathcal{O}(n)$, since $S(X)$ is obtained from $\phi$ by
      replacing each subformula in $M_F$ with a variable.
  \end{itemize}
  Overall, we have that the size of $\autom(\phi)$ is $\mathcal{O}(n)$.
\end{IEEEproof}

\begin{IEEEproof}
\label{app:proof:dssasize}
  Let $\phi$ be an \LTLEBR formula of size $n$. By \cref{prop:pastebrsize}, we
  can build an equivalent \PLTLEBR formula $\phi^\prime$ of size
  $\mathcal{O}(n^3 \cdot M^{\log_2 n + 1})$; by \cref{prop:canonizesize}, from
  $\phi^\prime$ we can obtain an equivalent canonical \PLTLEBR formula
  $\phi^{\prime\prime}$ of linear size with respect to $|\phi|$. Finally, by
  \cref{prop:dssasize}, the size of the deterministic symbolic safety automaton
  $\autom(\phi^{\prime\prime})$ is linear in $|\phi^\prime|$, hence
  $|\autom(\phi^{\prime\prime})| \in \mathcal{O}(n^3 \cdot M^{\log_2 n + 1})$.
\end{IEEEproof}

\begin{corollary}
\label{corol:dssasize}
  Let $\phi$ be an \LTLEBR formula with no constants, with $|\phi| = n$. Then, there exists a deterministic SSA of size $\mathcal{O}(n^3)$ that accepts the same language.
\end{corollary}
\begin{IEEEproof}
\label{app:proof:dssasize}
  Let $\phi$ be an \LTLEBR formula with no constants; then $M = 1$. By
  \cref{th:dssasize}, the size of the deterministic symbolic safety automaton
  recognizing the language of $\phi$ is $\mathcal{O}(n^3)$.
\end{IEEEproof}

\section{Plots}
\begin{figure}[h!]
  \centering
  \includegraphics[width=0.8\linewidth]{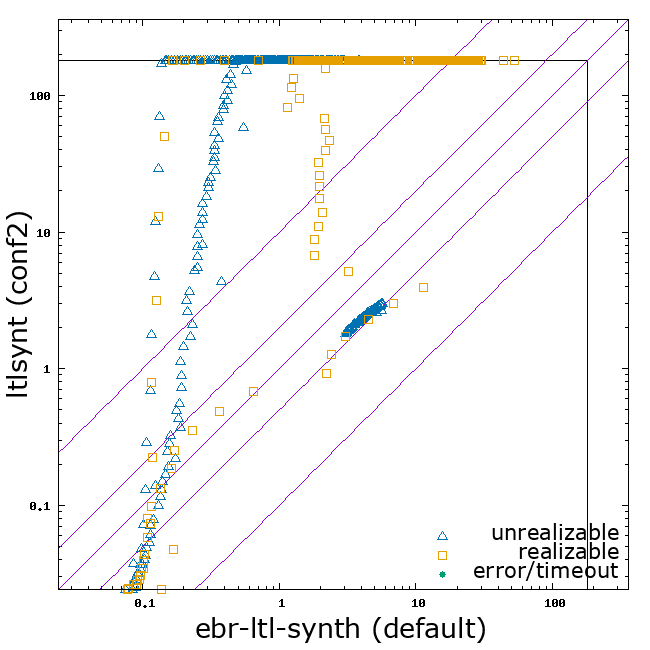}
  \caption{\tool vs \ltlsynt (second conf.) on all scalable benchmarks.}
  \label{fig:ebrltlsynt2}
\end{figure}%

\begin{figure}[h!]
  \centering
  \includegraphics[width=0.8\linewidth]{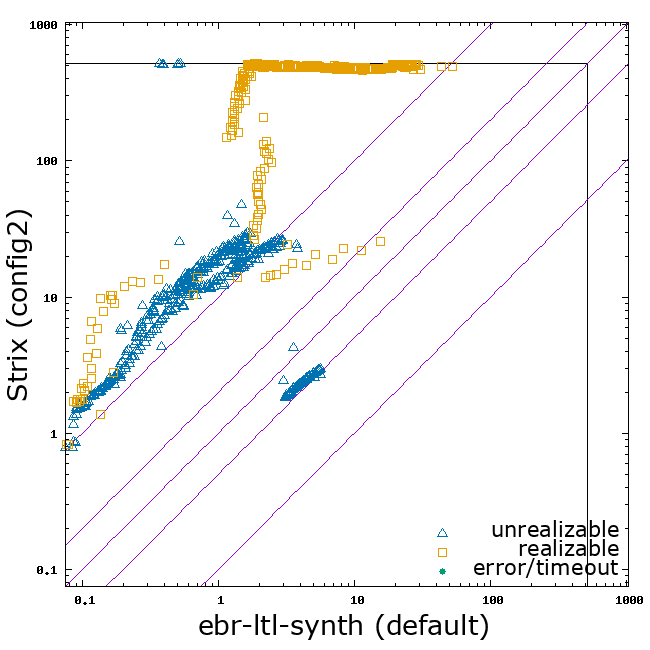}
  \caption{\tool vs \strix (second conf.) on all scalable benchmarks.}
  \label{fig:ebrstrix2}
\end{figure}%

\begin{figure}[h!]
  \centering
  \includegraphics[width=0.8\linewidth]{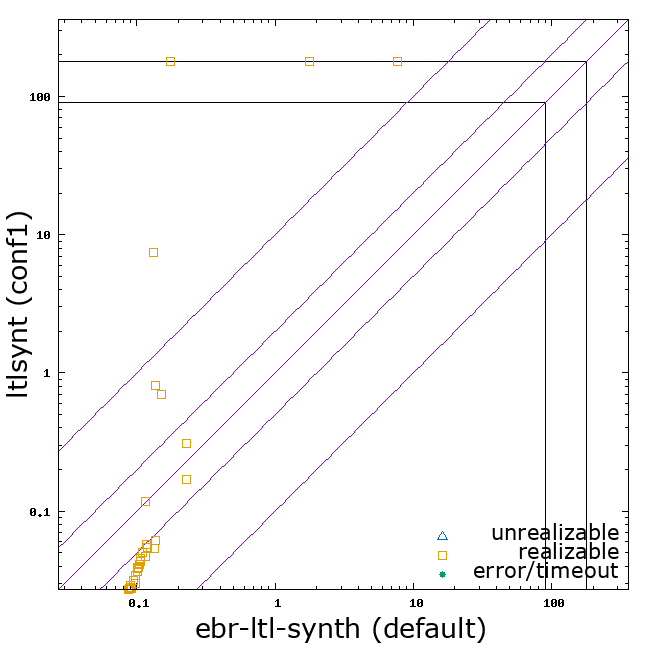}
  \caption{\tool vs \ltlsynt (first conf.) on SYNTCOMP benchmarks.}
\end{figure}%

\begin{figure}[h!]
  \centering
  \includegraphics[width=0.8\linewidth]{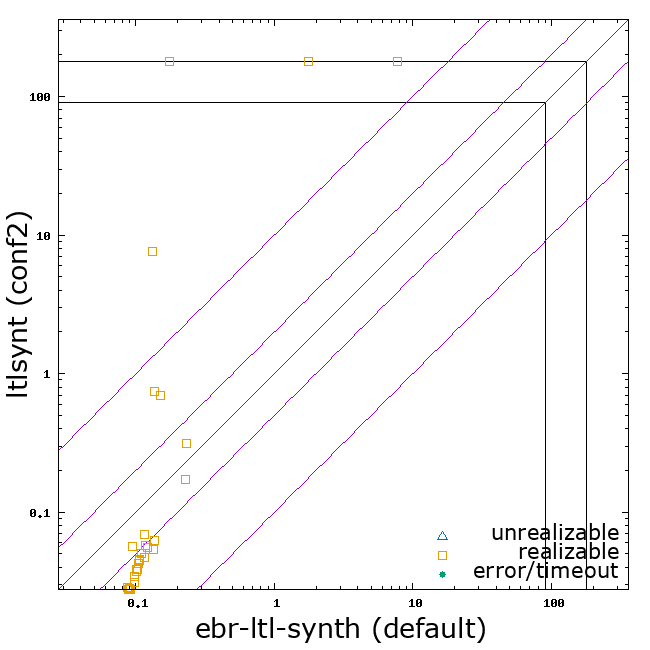}
  \caption{\tool vs \ltlsynt (second conf.) on SYNTCOMP benchmarks.}
\end{figure}%

\begin{figure}[h!]
  \centering
  \includegraphics[width=0.8\linewidth]{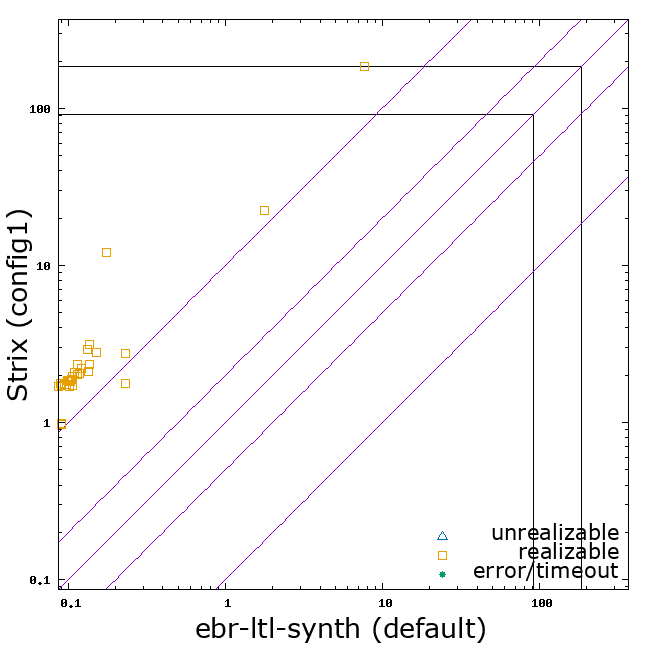}
  \caption{\tool vs \strix (first conf.) on SYNTCOMP benchmarks.}
\end{figure}%

\begin{figure}[h!]
  \centering
  \includegraphics[width=0.8\linewidth]{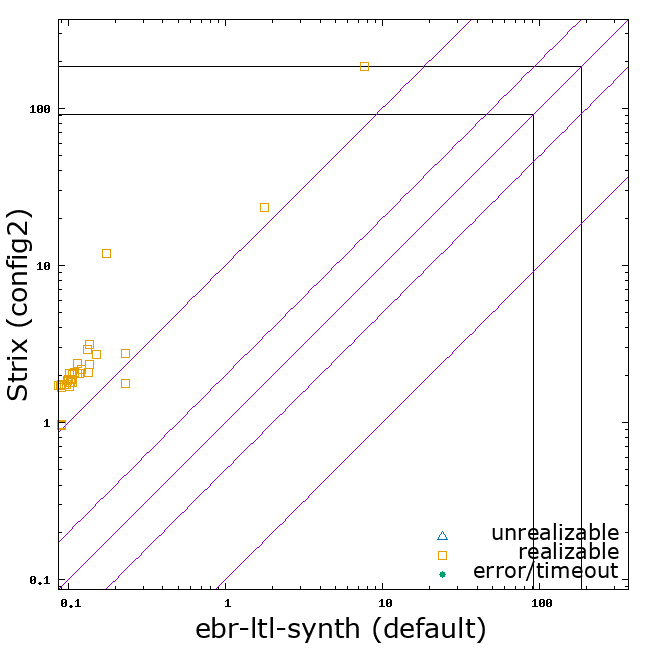}
  \caption{\tool vs \strix (second conf.) on SYNTCOMP benchmarks.}
\end{figure}%

\begin{figure}[h!]
  \centering
  \includegraphics[width=0.8\linewidth]{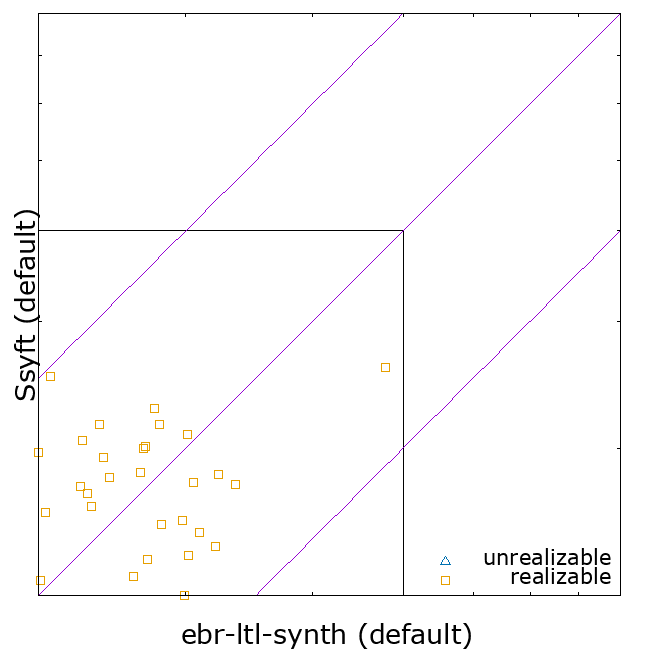}
  \caption{\tool vs \ssyft on SYNTCOMP benchmarks.}
\end{figure}%

\section{Pseudocodes}
\begin{figure}
\begin{lstlisting}[
  mathescape=true,
  frame=single,
  basicstyle=\scriptsize]{Name}
// Input: $\phi \in \LTLEBR$, in_future = false
// Output: $\phi \in \PLTLEBR$
$\toPastEbr$($\phi$, in_future){
  switch($\phi$){

    case $p$:
      return $p$;

    case $\lnot \phi_1$:
    case $\ltl{\psi_1 U^{[0,k]} \psi_2}$:
      return $\pastify(\phi)$
    
    case $\phi_1 \land \phi_2$:
      return $\toPastEbr$($\phi_1$, in_future) $\land$ 
             $\toPastEbr$($\phi_2$, in_future)

    case $\phi_1 \lor \phi_2$:
      if(in_future)
        return $\pastify(\phi)$
      else
        return $\toPastEbr$($\phi_1$, in_future) $\lor$
               $\toPastEbr$($\phi_2$, in_future)

    case $\ltl{X\phi_1}$:
      switch($\phi_1$){
        case $\phi_2 \land \phi_3$:
        case $\ltl{X\phi_2}$:
        case $\ltl{G\phi_2}$:
        case $\ltl{\psi R \phi_2}$:
          return $\ltl{X(}$$\toPastEbr$($\phi_1$, true)$)$
        default:
          return $\ltl{X(}\pastify(\phi_1))$
      }

    case $\ltl{G\phi_1}$:
      switch($\phi_1$){
        case $\phi_2 \land \phi_3$:
        case $\ltl{X\phi_2}$:
        case $\ltl{G\phi_2}$:
        case $\ltl{\psi R \phi_2}$:
          return $\ltl{G(}$$\toPastEbr$($\phi_1$, true)$)$
        default:
          return $\ltl{G(}\pastify(\phi_1))$
      }

    case $\ltl{\psi R \phi_1}$:
      switch($\phi_1$){
        case $\phi_2 \land \phi_3$:
        case $\ltl{X\phi_2}$:
        case $\ltl{G\phi_2}$:
        case $\ltl{\psi^\prime R \phi_2}$:
          return $\ltl{\psi R(}$$\toPastEbr$($\phi_1$, true)$)$
        default:
          return $\ltl{\psi R(}\pastify(\phi_1))$
      }
  }
}
\end{lstlisting}
  \caption{\toPastEbr algorithm.}
  \label{app:fig:pastify}
\end{figure}

\begin{figure}
\begin{lstlisting}[
  mathescape=true,
  frame=tlrb,
  basicstyle=\scriptsize]{Name}
// Input: $\phi \in \PLTLEBR$
// Output: $\phi \in$ canonical $\LTLEBR$
// Notation: 
//    $\phi,\phi_1,\dots,\phi_n \in \LTLEBR$
//    $\psi,\psi_1,\psi_2,\psi_3 \in \LTLFP$
//    $p \in \Sigma$
$\apply(\phi)${
  switch($\phi$){
    // Base case = $\LTLFP$ formulae
    case $p$ :
    case $\lnot \psi$ :
    case $\ltl{Y \psi_1}$ :
    case $\ltl{\psi_1 S \psi_2}$ :
      return $\phi$

    // And/Or Operators
    case $\phi_1 \land \phi_2$ :
    case $\phi_1 \lor \phi_2$ :
      return $\apply(\phi_1)$ $\land$ 
             $\apply(\phi_2)$

    // Next Rewriting Rules
    case $\ltl{X\phi_1}$ :
      $\phi_1 \gets \apply(\phi_1)$
      switch($\phi_1$){
        case $\phi_2 \land \dots \land \phi_n$ : // rule $R_1$
          return $\ltl{X\phi_2 \land \dots \land X\phi_n}$ 
        default : 
          return $\ltl{X\phi_1}$
      }

    // Globally Rewriting Rules
    case $\ltl{G\phi_1}$ :
      $\phi_1 \gets \apply(\phi_1)$
      switch($\phi_1$){
        case $\phi_2 \land \dots \land \phi_n$ : // rule $R_2$
          $\phi_2 \gets$ resolve_globally($\phi_2$)
          $\dots$
          $\phi_n \gets$ resolve_globally($\phi_n$)
          return $\phi_2 \land \dots \land \phi_n$
        default:
          $\phi_1 \gets$ resolve_globally($\phi_1$)
          return $\phi_1$
      }
      
    // Release Rewriting Rules
    case $\ltl{\psi R \phi_1}$ :
      $\phi_1 \gets \apply(\phi_1)$
      switch($\phi_1$){
        case $\phi_2 \land \dots \land \phi_n$ : // rule $R_2$ 
          $\phi_2 \gets$ resolve_release($\ltl{\psi}$,$\phi_2$)
          $\dots$
          $\phi_n \gets$ resolve_release($\ltl{\psi}$,$\phi_n$)
          return $\phi_2 \land \dots \land \phi_n$
        default : 
          $\phi_1 \gets$ resolve_release($\ltl{\psi}$,$\phi_1$)
          return $\phi_1$
      }

    default : 
      unreachable_code()
  }
}
\end{lstlisting}
  \caption{The $\apply$ algorithm (part I).}
  \label{app:fig:applyalg}
\end{figure}

\begin{figure}
\begin{lstlisting}[
  mathescape=true,
  frame=tlrb,
  basicstyle=\scriptsize]{Name}
resolve_globally($\phi$){
  switch($\phi$){
    case $\ltl{X^i \psi}$ : // rule $R_3$ (2nd case) 
      return $\ltl{X^iG\psi}$
    case $\ltl{X^iG\psi}$ : // rule $R_5$ 
      return $\ltl{X^iG\psi}$
    case $\ltl{X^i(\psi R \psi_1)}$ : // rule $R_6$ 
      return $\ltl{X^iG\psi_1}$
    default : 
      return $\ltl{G\psi}$
  }
}


resolve_release($\ltl{X^i\psi_1}$,$\phi$){
  switch($\phi$){
    case $\ltl{X^j \psi_2}$ : // rule $R_3$
      if($i>j$)
        return $\ltl{X^i(\psi_1 R (Y^{i-j}\psi_2))}$
      else
        return $\ltl{X^j((Y^{j-i}\psi_1) R \psi_2)}$
    case $\ltl{X^jG\psi_2}$ : // rule $R_7$ 
      if($i>j$)
        return $\ltl{X^iG(Y^{i-j}\psi_2)}$
      else
        return $\ltl{X^jG\psi_2}$
    case $\ltl{X^j(\psi_2 R \phi_3)}$ : // rule $R_4$
      if($i>j$)
        return $\ltl{X^i( \psi_1 R ((Y^{i-j}\psi_2) R (Y^{i-j} \psi_3) ) )}$
      else
        return $\ltl{X^j( (Y^{j-i}\psi_1 ) R (\psi_2 R \psi_3) )}$
    default : 
      return $\ltl{(X^i\psi_1) R \phi}$ 
  }
}
\end{lstlisting}
  \caption{The $\apply$ algorithm (part II).}
  \label{app:fig:resolve}
\end{figure}

\begin{figure}
\begin{lstlisting}[
  mathescape=true,
  frame=tlrb,
  basicstyle=\scriptsize]{Name}
$\flatten(\phi)${
  switch($\phi$){
    case $\phi_1 \land \phi_2$:
      return $\flatten(\phi_1) \land \flatten(\phi_2)$

    case $\phi_1 \lor \phi_2$:
      return $\flatten(\phi_1) \lor \flatten(\phi_2)$

    // rule $R_{flat}$
    case $\ltl{X^i( \psi_1 R (\psi_2 R (\dots (\psi_{n-1} R \psi_n) \dots)) )}$:
      return $\ltl{X^i( (\psi_{n-1} \land O(\psi_{n-2} \land \dots O(\psi_1
      \land Y^i \true)\dots)) R \psi_n )}$

    default:
      return $\phi$
  }
}
\end{lstlisting}
  \caption{The $\flatten$ algorithm.}
  \label{app:fig:flatten}
\end{figure}

\fi
\end{document}